\documentclass[aps,pra,twocolumn,10pt,showkeys,showpacs]{revtex4-1}

\usepackage{amsfonts,amsmath,amssymb}
\usepackage{mathtools}

\usepackage{url}

\allowdisplaybreaks[1]

\usepackage{hyperref} 

\usepackage{times}

\usepackage{bm}

\setcitestyle{numbers,square,aysep={},yysep={,},notesep={, },citesep={,}}
 
\usepackage{microtype}
 
\usepackage[capitalize]{cleveref}

\begin{document}

\title{Comment on \\
  ``Fourier transform of hydrogen-type atomic orbitals'', \\
  Can. J. Phys. Vol.\ 96, 724 - 726 (2018) \\
  by N. Y\"{u}k\c{c}\"{u} and S. A. Y\"{u}k\c{c}\"{u}}

\author{Ernst Joachim Weniger} 
\email{joachim.weniger@chemie.uni-regensburg.de,
  joachim.weniger@gmail.com}
\affiliation{Institut f\"{u}r
  Physikalische und Theoretische Chemie, Universit\"{a}t Regensburg,
  D-93040 Regensburg, Germany}

\begin{abstract}
  \citeauthor{Podolsky/Pauling/1929} (Phys. Rev. \textbf{34}, 109 - 116
  (1929)) were the first ones to derive an explicit expression for the
  Fourier transform of a bound-state hydrogen
  eigenfunction. \citeauthor{Yuekcue/Yuekcue/2018}, who were apparently
  unaware of the work of \citeauthor{Podolsky/Pauling/1929} or of the
  numerous other earlier references on this Fourier transform, proceeded
  differently. They expressed a generalized Laguerre polynomial as a
  finite sum of powers, or equivalently, they expressed a bound-state
  hydrogen eigenfunction as a finite sum of Slater-type functions. This
  approach looks very simple, but it leads to comparatively complicated
  expressions that cannot match the simplicity of the classic result
  obtained by \citeauthor{Podolsky/Pauling/1929}. It is, however,
  possible to reproduce not only the \citeauthor{Podolsky/Pauling/1929}
  formula for the bound-state hydrogen eigenfunction, but to obtain
  results of similar quality also for the Fourier transforms of other,
  closely related functions such as Sturmians, Lambda functions or
  Guseinov's functions by expanding generalized Laguerre polynomials in
  terms of so-called reduced Bessel functions.
\end{abstract}


\date{In Press, Canadian Journal of Physics, \today}%

\maketitle

%
\typeout{==> Section: Introduction}
\section{Introduction}
\label{Sec:Intro}
%

\citet*{Yuekcue/Yuekcue/2018} derived explicit expressions for the
Fourier transform of a bound-state hydrogen eigenfunction. Their article
creates the impression that their results \citep*[Eqs.\ (15) and
(16)]{Yuekcue/Yuekcue/2018} are new. This is wrong. Moreover, their
explicit expressions are less compact and therefore also less useful than
those already described in the literature.

In \citeyear{Podolsky/Pauling/1929}, \citet[Eq.\
(28)]{Podolsky/Pauling/1929} were the first ones to derive an explicit
expression via a direct Fourier transformation of a generating function
of the generalized Laguerre polynomials. In \citeyear{Hylleraas/1932},
\citet[Eqs.\ (11c) and (12)]{Hylleraas/1932} derived this Fourier
transformation algebraically by solving a differential equation for the
momentum space eigenfunction. In \citeyear{Fock/1935}, \citet{Fock/1935}
re-formulated the momentum space Schr\"{o}dinger equation for the
hydrogen atom as a 4-dimensional integral equation, whose solutions --
the 4-dimensional hyperspherical harmonics -- are nothing but the Fourier
transforms of bound-state hydrogen eigenfunctions in disguise (see for
example \citep[Section VI]{Weniger/1985} or the books by
\citet{Avery/1989,Avery/2000}, \citet{Avery/Avery/2006}, and
\citet*{Avery/Rettrup/Avery/2012} and references therein).

The Fourier transform of a bound state hydrogen eigenfunction has been
treated in numerous books and articles. Examples are the books by
\citet[Eq.\ (8.8)]{Bethe/Salpeter/1977}, \citet[Eqs.\ (5.5) and
(5.6)]{Englefield/1972}, or \citet*[Eq.\
(7.4.69)]{Biedenharn/Louck/1981a} or the relatively recent review by
\citet[Eq.\ (9.55)]{Hill/2006}. This Fourier transform was even discussed
in a Wikipedia article \citep{WikipediaHydAt/2018}, which cites the book
by \citet[Eq.\ (A5.34)]{Bransden/Joachain/1983} as its source. There are
also articles by \citet{Klein/1966} and by \citet{Hey/1993a,Hey/1993b},
which discuss properties of the momentum space hydrogen functions. In
\cite[Section IV]{Weniger/1985}, I presented a different and remarkably
simple derivation of the Fourier transform of a bound state hydrogen
eigenfunction and of related functions which will play a major role in
\cref{Sec:ExpRBF}.

In \cref{Sec:GLag_BoundStateHydrEigFun}, basic properties of the
generalized Laguerre polynomials and of bound state hydrogen
eigenfunctions are reviewed. As discussed in
\cref{Sec:Incom_BoundStateHydrEigFun}, bound state hydrogen
eigenfunctions are in contrast to several other similar function sets not
complete in the Hilbert space $L^{2} (\mathbb{R}^3)$ of square integrable
functions. This makes bound state hydrogen eigenfunctions useless in
expansions. Apparently, \citeauthor{Yuekcue/Yuekcue/2018} are unaware of
this well known and very consequential fact.

The explicit expressions derived by \citet*[Eqs.\ (15) and
(16)]{Yuekcue/Yuekcue/2018} are less useful than those mentioned above
(compare the discussion in \cref{Sec:WorkYuekcueYuekcue}). This is a
direct consequence of their derivation: \citeauthor{Yuekcue/Yuekcue/2018}
expressed generalized Laguerre polynomials in terms of
powers. Superficially, this looks quite natural, but actually it is a bad
idea. \Cref{Sec:ExpRBF} shows how their approach can be improved
substantially by expanding generalized Laguerre polynomials in terms of
better suited alternative function sets, the so-called reduced Bessel
functions.

%
\typeout{==> Section: Generalized Laguerre Polynomials and Bound-State
  Hydrogen Eigenfunctions}
\section{Generalized Laguerre Polynomials and Bound-State Hydrogen
  Eigenfunctions}
\label{Sec:GLag_BoundStateHydrEigFun}
%

The generalized Laguerre polynomials $L_{n}^{(\alpha)} (z)$ with
$\Re (\alpha) > - 1$ and $n \in \mathbb{N}_{0}$ are the classical
orthogonal polynomials associated with the integration interval
$[0, \infty)$ and the weight function $w (z) = z^{\alpha} \exp (-z)$.
They are of considerable importance in mathematics and also in
theoretical physics. There is a detailed literature which is far too
extensive to be cited here. Those interested in the historical
development with a special emphasis on quantum physics should consult an
article by \citet*{Mawhin/Ronveaux/2010}. Generalized Laguerre
polynomials also played a major role in my own research
\citep*{Weniger/Steinborn/1983a,Weniger/Steinborn/1984,Weniger/1985,%
  Weniger/2008,Weniger/2012,Borghi/Weniger/2015}.

It is recommendable to use the modern mathematical definition of the
generalized Laguerre polynomials $L_{n}^{(\alpha)} (z)$ with
$n \in \mathbb{N}_{0}$ and $\alpha, z \in \mathbb{C}$, which are defined
either via their Rodrigues' relationship \citep*[Eq.\ (18.5.5) and Table
18.5.1]{Olver/Lozier/Boisvert/Clark/2010} or as a terminating confluent
hypergeometric series ${}_{1} F_{1}$ \citep*[Eq.\
(18.5.12)]{Olver/Lozier/Boisvert/Clark/2010}:
\begin{align}
  \label{GLag_Rodrigues}
  L_{n}^{(\alpha)} (z) & \; = \; z^{-\alpha} \, 
   \frac{\mathrm{e}^{z}}{n!} \, \frac{\mathrm{d}^n}{\mathrm{d} z^n} \, 
    \bigl[ \mathrm{e}^{-z} z^{n+\alpha} \bigr] 
  \\
  \label{GLag_1F1}
  & \; = \; \frac{(\alpha+1)_n}{n!} \, {}_1 F_1 (-n; \alpha+1; z) \, .
\end{align}
Further details can be found in books on special functions.

Dating back from the early days of quantum mechanics, an antiquated
notation is still frequently used mainly in atomic theory. For example,
\citet*[Eq.\ (3.5)]{Bethe/Salpeter/1977} introduced so-called
\emph{associated Laguerre functions}
$\bigl[ L_{n}^{m} (z) \bigr]_{\text{BS}}$ with $n, m \in \mathbb{N}_0$
via the Rodrigues-type relationships
\begin{subequations}
  \label{AssLagFun_BS}
  \begin{align}
    \label{AssLagFun_BS_1}
    \bigl[ L_{n}^{m} (z) \bigr]_{\text{BS}} & \; = \;
     \frac{\mathrm{d}^{m}}{\mathrm{d} z^{m}} \, 
      \bigl[L_{n} (z) \bigr]_{\text{BS}} \, ,
    \\
    \label{AssLagFun_BS_2}
    \bigl[ L_{n} (z) \bigr]_{\text{BS}} & \; = \; \mathrm{e}^{z} \,
     \frac{\mathrm{d}^{n}}{\mathrm{d} z^{n}} \, \bigl[ \mathrm{e}^{-z}
      z^{n} \bigr] \, .
  \end{align} 
\end{subequations}
This convention is also used in the books by \citet*[Eqs.\ (6) and (9) on
p.\ 115]{Condon/Shortley/1970} and \citet*[Eq.\ (2) on p.\
189]{Condon/Odabasi/1980}.

Generalized Laguerre polynomials with integral superscript
$\alpha=m \in \mathbb{N}_{0}$ and the associated Laguerre functions
\eqref{AssLagFun_BS} are connected via
\begin{equation}
  \label{GLagPol_2_GLagPol_BS}
  L_{n}^{(m)} (z) \; = \; \frac{(-1)^m}{(n+m)!} \, 
   \bigl[ L_{n+m}^{m} (z) \bigr]_{\text{BS}} \, ,
    \quad m, n \in \mathbb{N}_{0} \, .
\end{equation}
The notation for associated Laguerre functions is less intuitive than the
notation for the generalized Laguerre polynomials, whose subscript $n$
corresponds to the polynomial degree and whose superscript $\alpha$
characterizes the weight function $w (z) = z^{\alpha} \exp (-z)$. The
worst drawback of the functions \eqref{AssLagFun_BS} is that they cannot
express generalized Laguerre polynomials $L_{n}^{(\alpha)}$ with
\emph{non-integral} superscripts $\alpha$ which also occur in quantum
physics. The eigenfunctions $\Omega_{n, \ell}^{m} (\beta, \bm{r})$ of the
Hamiltonian $\beta^{-2} \nabla^2 - \beta^2 r^2$ of the three-dimensional
isotropic harmonic oscillator contain generalized Laguerre polynomials in
$r^{2}$ with half-integral superscripts (see for example \citep[Eq.\
(5.4)]{Weniger/1985} and references therein). Similarly, the
eigenfunctions of the Dirac equation for the hydrogen atom contain
generalized Laguerre polynomials with in general non-integral
superscripts \citep[Eqs.\ (9.84) and (9.85)]{Hill/2006}.

If the modern mathematical notation is used, the bound-state
eigenfunctions of a hydrogenlike ion with nuclear charge $Z$ in spherical
polar coordinates is essentially the product of an exponential and a
generalized Laguerre polynomial, both depending on $r$, and a regular
solid harmonic
$\mathcal{Y}_{\ell}^{m} (\bm{r}) = r^{\ell} Y_{\ell}^{m} (\theta, \phi)$
(see for example \citep*[Eqs.\ (7.4.41) -
(7.4.43)]{Biedenharn/Louck/1981a} or \citep[Eqs.\ (9.2) and
(9.10)]{Hill/2006}):
\begin{align}
  \label{Def_HydEigFun}
  & W_{n, \ell}^{m} (Z, \bm{r}) \; = \; \left( \frac{2Z}{n} \right)^{3/2}
   \, \left[ \frac{(n-\ell-1)!}{2n(n+\ell)!} \right]^{1/2}
  \notag 
  \\
  & \qquad \times \, \mathrm{e}^{-Zr/n} \, L_{n-\ell-1}^{(2\ell+1)}
   (2Zr/n) \, \mathcal{Y}_{\ell}^{m} (2Z \bm{r}/n) \, ,
  \notag 
  \\
  & \qquad \quad \; n \in \mathbb{N} \, , \; 
   \ell \in \mathbb{N}_0 \le n-1 \, , \; -\ell \le m \le \ell \, .
\end{align}

\citet{Yuekcue/Yuekcue/2018} define the radial part of the bound-state
eigenfunctions \eqref{Def_HydEigFun} via their Eq.\ (3), which is
inconsistent with their definition of the generalized Laguerre
polynomials via their Eq.\ (11). It can be shown that their Eq.\ (11) is
equivalent to \cref{GLag_1F1} which implies that
\citeauthor{Yuekcue/Yuekcue/2018} also use the modern mathematical
notation. In addition, their Ref.\ [26] for their Eq.\ (11) is
incorrect. The so-called \emph{Bateman Manuscript Project}
\citep{Erdelyi/Magnus/Oberhettinger/Tricomi/HTF1/1953a,%
  Erdelyi/Magnus/Oberhettinger/Tricomi/HTF2/1953b,%
  Erdelyi/Magnus/Oberhettinger/Tricomi/HTF3/1953c,%
  Erdelyi/Magnus/Oberhettinger/Tricomi/TIT1/1954a,%
  Erdelyi/Magnus/Oberhettinger/Tricomi/TIT2/1954b} was named to honor
Harry Bateman who had died in 1946, i.e., long before these books had
been completed. Thus, the correct reference for Eq.\ (11) of
\citet{Yuekcue/Yuekcue/2018} would be \citep[Eq.\ (7) on p.\
188]{Erdelyi/Magnus/Oberhettinger/Tricomi/HTF2/1953b}.

%
\typeout{==> Section: Incompleteness of the Bound-State Hydrogen
  Eigenfunctions}
\section{Incompleteness of the Bound-State Hydrogen Eigenfunctions}
\label{Sec:Incom_BoundStateHydrEigFun}
%

Expansions of a given function in terms of suitable function sets are
among the most useful techniques of mathematical physics. This approach
requires that the function set being used is complete and preferably also
orthogonal in the corresponding Hilbert space. As for example discussed
in \citep{Weniger/2012} or in \citep{Klahn/1981}, non-orthogonal
expansions can easily have pathological properties.

Bound-state hydrogenic eigenfunctions \eqref{Def_HydEigFun} are
orthonormal with respect to an integration over the whole $\mathbb{R}^3$,
\begin{equation}
  \int \, \bigl[ W_{n, \ell}^{m} (Z, \bm{r}) \bigr]^{*} \,  
  W_{n', \ell'}^{m'} (Z, \bm{r}) \, \mathrm{d}^3 \bm{r} 
  \; = \; \delta_{n n'} \, \delta_{\ell \ell'} \, \delta_{m m'} \, ,
\end{equation}
but they are not complete in the Hilbert space 
\begin{equation}
  \label{HilbertL^2}
  L^{2} (\mathbb{R}^3) \; = \; \Bigl\{ f \colon \mathbb{R}^3 \to
   \mathbb{C} \Bigm\vert \, \int \, \vert f (\bm{r}) \vert^2 \,
    \mathrm{d}^3 \bm{r} < \infty \Bigr\}
\end{equation}
of square integrable functions \emph{without} the inclusion of the
technically very difficult continuum eigenfunctions, described for
instance in \citep[pp.\ 21 - 25]{Bethe/Salpeter/1977}, in \citep*[Chapter
33 Coulomb Functions]{Olver/Lozier/Boisvert/Clark/2010} or in the recent
article \citep{Gaspard/2018}. \citeauthor{Yuekcue/Yuekcue/2018} are
apparently not aware of this incompleteness.

In the literature, this incompleteness, which was first described in
\citeyear{Hylleraas/1928} by \citet[p.\ 469]{Hylleraas/1928}, is
sometimes overlooked -- often with catastrophic consequences. For
example, \citeauthor{Yuekcue/Yuekcue/2018} cited as their Ref.\ [4] an
article by \citet{Yamaguchi/1983} in order to demonstrate the usefulness
of bound-state hydrogen eigenfunctions in expansions. However,
\citeauthor{Yamaguchi/1983}'s article had been severely criticized in
\citep{Weniger/Steinborn/1984} for simply neglecting the troublesome
continuum eigenfunctions. Already in 1955, \citet{Shull/Loewdin/1955} had
emphasized the importance of the continuum eigenfunctions and tried to
estimate the magnitude of the error due to their omission.

At first sight, this incompleteness may seem surprising since the
completeness of the generalized Laguerre polynomials
$L_{n}^{(\alpha)} (z)$ in the weighted Hilbert space
\begin{align}
  \label{HilbertL^2_Lag}
  & L^{2}_{\mathrm{e}^{-z} z^{\alpha}} \bigl([0, \infty) \bigr)
  \notag
  \\
  & \quad \; = \;
   \Bigl\{ f \colon \mathbb{C} \to \mathbb{C} \Bigm\vert \,
    \int_{0}^{\infty} \,\mathrm{e}^{-z} \, z^{\alpha} \, \vert f (z)
     \vert^2 \, \mathrm{d} z < \infty \Bigr\}  
\end{align}
is a classic result of mathematical analysis (see for example the books
by \citet[p.\ 33]{Higgins/1977}, \citet[pp.\ 349 - 351]{Sansone/1977},
\citet[pp.\ 108 - 110]{Szegoe/1975}, or \citet[pp.\ 235 -
238]{Tricomi/1970}). Thus, every function
$f \in L^{2}_{\mathrm{e}^{-z} z^{\alpha}} \bigl([0, \infty) \bigr)$ can
be expressed by a Laguerre series
\begin{subequations} 
  \label{f_Exp_GLag}
  \begin{align}
    \label{f_Exp_GLag_a}
    f (z) \; = \; & \sum_{n=0}^{\infty} \,
     \lambda_{n}^{(\alpha)} \, L_{n}^{(\alpha)} (z) \, ,
    \\
    \label{f_Exp_GLag_b}
    \lambda_{n}^{(\alpha)} \; = \; & \frac{n!}{\Gamma (\alpha+n+1)} \,
     \int_{0}^{\infty} \, z^{\alpha} \, \mathrm{e}^{-z} \, 
      L_{n}^{(\alpha)} (z) \, f (z) \, \mathrm{d} z \, ,
  \end{align}
\end{subequations}
which converges in the mean with respect to the norm of the Hilbert space
$L^{2}_{\mathrm{e}^{-z} z^{\alpha}} \bigl([0, \infty) \bigr)$. For a
condensed discussion of Laguerre expansions, see \citep[Section
2]{Weniger/2008}.

How can the incompleteness of the bound-state hydrogen eigenfunctions
\eqref{Def_HydEigFun} be explained? The culprit is their $n$-dependent
scaling parameter $2Z/n$. \citet[Eq.\ (6.17) on p.\ 200]{Fock/1978} showed
that the confluent hypergeometric function
\begin{align}
  \label{1F1_HydEigFun}
  & {}_{1} F_{1} \bigl( -n+\ell+1; 2\ell + 2; 2 Z r/n \bigr)
  \notag 
  \\
  & \quad \; = \;
   \sum_{\nu=0}^{n-\ell-1} \, 
    \frac{(-n+\ell+1)_{\nu}}{(2\ell+2)_{\nu}} \, 
     \frac{[2Z r/n]^{\nu}}{\nu!}
\end{align}
occurring in \cref{Def_HydEigFun} can in the limit $n \to \infty$ be
represented by a Bessel function
$J_{2\ell+1} \left( \sqrt{8 Z r} \right)$ of the first kind, which is an
oscillatory function that decays too slowly to be square integrable
(compare also \citep*[Eq.\
(18.11.6)]{Olver/Lozier/Boisvert/Clark/2010}). In the limit
$n \to \infty$, the exponential $\exp (-Z r/n)$ in \cref{Def_HydEigFun}
loses its exponential decay as $r \to \infty$. Consequently, the bound
state hydrogen eigenfunctions \eqref{Def_HydEigFun} become oscillatory as
$n \to \infty$, which means that they are no longer square
integrable. Instead, they belong to the continuous spectrum. Thus, the
so-called bound-state eigenfunctions are no longer bound-state functions
if the principal quantum number $n$ becomes very large. This implies that
the bound-state eigenfunctions cannot form a basis for the Hilbert space
$L^{2} (\mathbb{R}^3)$ of square integrable functions (compare
\citep[text following Eq.\ (6.19) on p.\ 201]{Fock/1978}).

Because of the incompleteness of the bound-state hydrogen eigenfunction,
it is now common to use in expansions alternative function sets also
based on the generalized Laguerre polynomials that possess more
convenient completeness properties. Closely related to the bound-state
hydrogenic eigenfunctions are the so-called Coulomb Sturmians or
Sturmians which were already used in \citeyear{Hylleraas/1928} by
\citet[Eq.\ (25) on p.\ 478]{Hylleraas/1928}:
\begin{align}
  \label{Def_SturmFun}
  & \Psi_{n, \ell}^{m} (\beta, \bm{r}) \; = \; (2 \beta)^{3/2} \, 
   \left[ \frac{(n-\ell-1)!}{2n(n+\ell)!} \right]^{1/2}
  \notag 
  \\
   & \qquad \, \times \, 
     \mathrm{e}^{-\beta r}  \, L_{n-\ell-1}^{(2\ell+1)} (2\beta r) \,
      \mathcal{Y}_{\ell}^{m} (2 \beta \bm{r}) \, .  
\end{align}
Here, the notation of \citep[Eq.\ (4.6)]{Weniger/1985} is used. We obtain
bound-state hydrogen eigenfunctions \eqref{Def_HydEigFun} with a correct
normalization factor if we make in \cref{Def_SturmFun} the substitution
$\beta \mapsto Z/n$ (compare the discussion following \citep[Eq.\
(4.12)]{Weniger/1985}):
\begin{equation}
  \label{SturmFun<->BSHEF}
  \Psi_{n, \ell}^{m} (Z/n, \bm{r}) \; = \; 
   W_{n, \ell}^{m} (Z, \bm{r}) \, .
\end{equation}
This is a non-trivial result. Sturmians are complete and orthonormal in
the in the Sobolev space $W_{2}^{(1)} (\mathbb{R}^3)$ (for the definition
of Sobolev spaces plus further references, see \citep[Section
II]{Weniger/1985}), whereas bound state hydrogen functions are
orthonormal but incomplete in the Hilbert space $L^{2} (\mathbb{R}^3)$.

Sturmians occur in the context of \citeauthor{Fock/1935}'s treatment of
the hydrogen atom \citep{Fock/1935}, albeit in a somewhat disguised form
(compare \citep[Section VI]{Weniger/1985}). There is a classic review by
\citet{Rotenberg/1970}. A fairly detailed discussion of their properties
was given by \citet{Novosadov/1983}. Sturmians also play a major role in
books by \citet{Avery/1989,Avery/2000}, \citet{Avery/Avery/2006}, and
\citet*{Avery/Rettrup/Avery/2012}. We used Sturmians for the construction
for an addition theorem of the Yukawa potential
\citep*{Homeier/Weniger/Steinborn/1992a} with the help of weakly
convergent orthogonal and biorthogonal expansions for the plane wave
introduced in \citep[Section III]{Weniger/1985}.

Lambda functions were introduced already in \citeyear{Hylleraas/1929} by
\citet[Footnote~${}^{*}$ on p.\ 349]{Hylleraas/1929}, and later by
\citet{Shull/Loewdin/1955} and by \citet[Eq.\ (46)]{Loewdin/Shull/1956}:
\begin{align}
  \label{Def_LambdaFun}
  & \Lambda_{n, \ell}^{m} (\beta, \bm{r}) \; = \; (2 \beta)^{3/2}
   \left[ \frac{(n-\ell-1)!} {(n+\ell+1)!} \right]^{1/2}
  \notag
  \\
  & \qquad \times \,
   \mathrm{e}^{-\beta r} \, L_{n-\ell-1}^{(2\ell+2)} (2 \beta r) \,
    \mathcal{Y}_{\ell}^{m} (2 \beta \bm{r}) \, .
\end{align}
Here, the notation of \citep[Eq.\ (4.4)]{Weniger/1985} is used.

The use of Lambda functions in electronic structure theory was suggested
by \citet{Kutzelnigg/1963} and \citet{Smeyers/1966} in
\citeyear{Kutzelnigg/1963} and \citeyear{Smeyers/1966},
respectively. \citet{Filter/Steinborn/1980} used them for the derivation
of one-range addition theorems of exponentially decaying functions, and I
used both Sturmians and Lambda functions for the construction of weakly
convergent expansions of a plane wave \citep{Weniger/1985}.

Both Sturmians and Lambda functions defined by
\cref{Def_SturmFun,Def_LambdaFun} have a fixed scaling parameter
$\beta > 0$ that does not depend on the principal quantum number
$n$. Consequently, these functions are orthogonal and complete in
suitable Hilbert and Sobolev spaces. A detailed discussion of the
mathematical properties of the functions
$\Psi_{n, \ell}^{m} (\beta, \bm{r})$ and
$\Lambda_{n, \ell}^{m} (\beta, \bm{r})$ was given in \citep[Section
IV]{Weniger/1985} or in \citep[Section 2]{Weniger/2012}.

%
\typeout{==> Section: The Work of Podolsky and Pauling}
\section{The Work of Podolsky and Pauling}
\label{Sec:PodolskyPauling}
%

The Fourier transform of an irreducible spherical tensor of integral rank
yields a Hankel-type radial integral multiplied by a spherical harmonic
if the so-called Rayleigh expansion of a plane wave (compare for instance
\citep*[p.\ 442]{Biedenharn/Louck/1981a}) is used:
\begin{align}
  \label{Rayleigh_Expan}
  & \mathrm{e}^{\pm \mathrm{i} \bm{x} \cdot \bm{y}} \; = \;
   4\pi \, \sum_{\ell=0}^{\infty} \, (\pm \mathrm{i})^{\ell} \, 
    j_{\ell} (xy) 
  \notag
  \\
  & \quad \times \,  \sum_{m=-\ell}^{\ell} \, 
   \bigl[ Y_{\ell}^{m} (\bm{x}/x) \bigr]^{*} \, Y_{\ell}^{m} (\bm{y}/y) 
    \, , \quad \bm{x}, \bm{y} \in \mathbb{R}^{3} \, .
\end{align}
With the help of the orthonormality of the spherical harmonics and the
definition of the spherical Bessel functions $j_{\ell} (xy)$ (see for
example \citep*[Eq.\ (10.47.3)]{Olver/Lozier/Boisvert/Clark/2010}), we
obtain the following expression for the Fourier transformation of a
Sturmian function \eqref{Def_SturmFun} without normalization factor and
with fixed $\beta > 0$:
\begin{align}
  \label{FT_Sturm_1}
  & (2 \pi)^{-3/2} \, 
   \int \, \mathrm{e}^{\mathrm{i} \bm{p} \cdot \bm{r}} \, 
    \mathrm{e}^{-\beta r} \, r^{\ell} \, L_{n}^{(2\ell+1)} (2 \beta r)
     \, Y_{\ell}^{m} (\bm{r}/r) \, \mathrm{d}^{3} \bm{r} 
  \notag
  \\
  & \; \; = \; (-\mathrm{i})^{\ell} \, p^{-1/2} \, 
   Y_{\ell}^{m} (\bm{p}/p)  
  \notag
  \\
  & \quad \times \, \int_{0}^{\infty} \, r^{\ell+3/2} \, 
    \mathrm{e}^{-\beta r} \, J_{\ell+1/2} (p r) \, 
     L_{n}^{(2\ell+1)} (2 \beta r) \, \mathrm{d} r \, .   
\end{align}
For a closed form expression of the Hankel-type radial integral in
\cref{FT_Sturm_1}, we need an explicit expression for the integral
\begin{equation}
  \label{HankelInt_GLag}
  I_{n}^{(\alpha, \mu, \nu)} (a, b) \; = \; \int_{0}^{\infty} \, y^{\mu}
   \, \mathrm{e}^{-ay} \, J_{\nu} (by) \, L_{n}^{(\alpha)} (2ay) \, 
    \mathrm{d} y \, .
\end{equation}
In \citeyear{Podolsky/Pauling/1929}, when \citet*{Podolsky/Pauling/1929}
tried to derive an expression for the Fourier transform of a bound-state
hydrogen eigenfunction, no explicit expression for this integral was
known. Even today, I could not find the required expression in the usual
books on special function theory.

\citet[Eq.\ (6)]{Podolsky/Pauling/1929} found a very elegant solution to
this problem. Their starting point was the generating function
\citep*[p.\ 242]{Magnus/Oberhettinger/Soni/1966}
\begin{equation}
  \label{GLag_GenFun}
  \frac {\exp \left( \frac{xt}{t-1} \right)}{(1-t)^{\alpha+1}} \; = \;
   \sum_{n=0}^{\infty} \, L_{n}^{(\alpha)} (x) \, t^{n} \, ,
    \quad \vert t \vert < 1 \, .    
\end{equation} 
Inserting this generating function of the generalized Laguerre
polynomials into the radial integral in \cref{FT_Sturm_1} yields:
\begin{align}
  \label{FT_Sturm_2}
  & \sum_{n=0}^{\infty} \, t^{n} \, 
   \int_{0}^{\infty} \, r^{\ell+3/2} \, 
    \mathrm{e}^{-\beta r} \, J_{\ell+1/2} (p r) \, 
     L_{n}^{(2\ell+1)} (2 \beta r) \, \mathrm{d} r 
  \notag
  \\  
  & \quad \; = \; (1-t)^{-2\ell-2} 
  \notag
  \\ 
  & \qquad \times \int_{0}^{\infty} \, 
   \mathrm{e}^{-\beta r \frac{1+t}{1-t}} \,
    \, r^{\ell+3/2} J_{\ell+1/2} (p r) \, \mathrm{d} r \, .  
\end{align}

The radial integral on the right-hand side can be expressed in closed
form. We use \citep[Eq.\ (2) on p.\ 385]{Watson/1922} 
\begin{align}
  \label{Int_ExpBesJ}
  & \int_{0}^{\infty} \, \mathrm{e}^{-ay} \, J_{\nu} (by) \, y^{\mu-1}
   \, \mathrm{d} y \; = \; \frac{(b/2)^{\nu} \Gamma (\mu+\nu)}
    {a^{\mu+\nu} \Gamma (\nu+1)} 
  \notag 
  \\
  & \qquad \times \, {}_{2} F_{1} \left( \frac{\mu+\nu}{2},
   \frac{\mu+\nu+1}{2}; \nu+1; - \frac{b^2}{a^2} \right) \, ,
  \notag
  \\
  & \qquad \qquad \Re (a \pm \mathrm{i} b) > 0 \, ,
\end{align}
to obtain
\begin{align}
  \label{FT_Sturm_4}
  & \int_{0}^{\infty} \, \mathrm{e}^{-\beta r \frac{1+t}{1-t}} \,
   \, r^{\ell+3/2} J_{\ell+1/2} (p r) \, \mathrm{d} r
  \notag
  \\
  & \; \; = \; \frac {(2\ell+2)!} {\Gamma (\ell+3/2)} \, \frac
   {(p/2)^{\ell+1/2}} 
    {\bigl[ \beta (1+t)/(1-t) \bigr]^{2\ell+3}}
  \notag
  \\[1\jot]
  & \quad \
   \times {}_{2} F_{1} \left( \ell+3/2, \ell+2; \ell+3/2; 
    - \frac{p^{2}}{\beta^{2}} \frac{(1-t)^{2}}{(1+t)^{2}} \right) \, .
\end{align}
This Gaussian hypergeometric series ${}_{2} F_{1}$ is actually a binomial
series
${}_{1} F_0 (\ell+2; z) = \sum_{m=0}^{\infty} (\ell+2)_{m} z^{m} / m! =
(1-z)^{-\ell -2}$
with $z = - p^{2} (1-t)^{2}/[\beta^{2} [1+t]^{2}]$ \citep*[Eq.\
(15.4.6)]{Olver/Lozier/Boisvert/Clark/2010}. Thus, we obtain for the
right-hand side of \cref{FT_Sturm_2}: 
\begin{align}
  \label{FT_Sturm_6}
  & (1-t)^{-2\ell-2} \, 
   \int_{0}^{\infty} \, \mathrm{e}^{-\beta r \frac{1+t}{1-t}} \,
    \, r^{\ell+3/2} J_{\ell+1/2} (p r) \, \mathrm{d} r 
  \notag
  \\
  & \quad \; = \; \frac
   {(2\ell+2)!} {\Gamma (\ell+3/2)} \,
    \frac {(p/2)^{\ell+1/2} \beta (1-t^{2})} 
     {\bigl[ \beta^{2} (1+t)^{2} + p^{2} (1-t)^{2} \bigr]^{\ell+2}}
\end{align} 
The denominator can be simplified further, using
$\beta^{2} (1+t)^{2} + p^{2} (1-t)^{2} = \bigl( \beta^{2} + p^{2} \bigr)
\bigl\{ 1 + \bigl[2 \bigl( \beta^{2} - p^{2} \bigr) \bigr] \ \bigl[
\beta^{2} + p^{2} \bigr] t + t^{2} \bigr\}$, yielding
\begin{align}
  \label{FT_Sturm_9}
  & (1-t)^{-2\ell-2} \, 
   \int_{0}^{\infty} \, \mathrm{e}^{-\beta r \frac{1+t}{1-t}} \,
    \, r^{\ell+3/2} J_{\ell+1/2} (p r) \, \mathrm{d} r 
  \notag
  \\
  & \quad \; = \; \frac
   {(2\ell+2)!} {\Gamma (\ell+3/2)} \, 
  \notag
  \\
  & \qquad \times \,
   \frac {(p/2)^{\ell+1/2} \beta (1-t^{2})} 
    {\displaystyle \left[ \bigl( \beta^{2} + p^{2} \bigr) 
     \left\{ 1 + \frac {2 \bigl( \beta^{2} - p^{2} \bigr)}
      {\beta^{2} + p^{2}} t + t^{2} \right\} \right]^{\ell+2}} \, .
\end{align}
The rational function on the right-hand side closely resembles the
generating function \citep*[p.\ 222]{Magnus/Oberhettinger/Soni/1966}
\begin{equation}
  \label{GegPol_GenFun}
  \bigl( 1 - 2 x t + t^{2} \bigr)^{-\lambda} \; = \;
   \sum_{n=0}^{\infty} \, C_{n}^{\lambda} (x) \, t^{n} \, , 
    \quad \vert t \vert < 1 \, ,
\end{equation}
of the Gegenbauer polynomials. \citeauthor{Podolsky/Pauling/1929} only
had apply the differential operator
$t^{1-\lambda} [\partial/\partial t] t^{\lambda}$ to
\cref{GegPol_GenFun}. This yields the following modified generating
function of the Gegenbauer polynomials (compare \citep*[Eq.\ (25
)]{Podolsky/Pauling/1929}),
\begin{equation}
  \label{Mod_GegPol_GenFun}
  \frac {1-t^{2}} {\bigl( 1 - 2 x t + t^{2} \bigr)^{\lambda+1}} 
   \; = \; \sum_{n=0}^{\infty} \, \frac{\lambda+n}{\lambda} \,
    C_{n}^{\lambda} (x) \, t^{n} \, ,
\end{equation}
which I could not find in the usual books on special function theory. The
rational function on the right-hand side of \cref{FT_Sturm_9} is of the
same type as the left-hand side of this modified generating function. If
we make in \cref{Mod_GegPol_GenFun} the substitutions
$x \mapsto (p^{2}-\beta^{2})/(p^{2}+\beta^{2})$ and
$\lambda \mapsto \ell+1$, we obtain the following expansion in terms of
Gegenbauer polynomials:
\begin{align}
  \label{FT_Sturm_10}
  & \bigl(1-t^{2}\bigr) \bigg/ {\displaystyle 
   \left\{ 1 - 2 \frac {p^{2}-\beta^{2}}
    {p^{2}+\beta^{2}}t + t^{2} \right\}^{\ell+2}}
  \notag
  \\
  & \qquad \; = \;
     \sum_{n=0}^{\infty} \, \frac{n+\ell+1}{\ell+1} \, 
      C_{n}^{\ell+1} \left( \frac {p^{2}-\beta^{2}}
    {p^{2}+\beta^{2}} \right) \, t^{n} \, .  
\end{align}
Inserting this into \cref{FT_Sturm_9} yields:
\begin{align}
  \label{FT_Sturm_11}
    & (1-t)^{-2\ell-2} \, 
   \int_{0}^{\infty} \, \mathrm{e}^{-\beta r \frac{1+t}{1-t}} \,
    \, r^{\ell+3/2} J_{\ell+1/2} (p r) \, \mathrm{d} r 
  \notag
  \\
  & \quad \; = \; \frac
   {(p/2)^{\ell+1/2} (2\ell+2)!} {(\ell+1) \Gamma (\ell+3/2)} \, 
    \beta 
  \notag
  \\
  & \qquad \times \, \sum_{n=0}^{\infty} \, \frac {n+\ell+1}
   {\bigl[ p^{2}+\beta^{2} \bigr]^{\ell+2}} \, 
    C_{n}^{\ell+1} \left( \frac {p^{2}-\beta^{2}}
     {p^{2}+\beta^{2}} \right) \, t^{n} \, .
\end{align}
Thus, we finally obtain the following explicit expression for the Fourier
transform of an unnormalized Sturmian:
\begin{align}
  \label{FT_Sturm_13}
  & (2 \pi)^{-3/2} \, 
   \int \, \mathrm{e}^{\mathrm{i} \bm{p} \cdot \bm{r}} \, 
    \mathrm{e}^{-\beta r} \, L_{n-\ell-1}^{(2\ell+1)} (2 \beta r)
     \, \mathcal{Y}_{\ell}^{m} (2 \beta \bm{r}) \, \mathrm{d}^{3} \bm{r}
  \notag
  \\
  & \quad \; = \; \frac 
   {(2/\pi)^{1/2} \, 2^{2\ell+1} \, \ell! \, \beta^{\ell+1} \, n}
    {\bigl[ p^{2}+\beta^{2} \bigr]^{\ell+2}} 
  \notag
  \\
  & \qquad \times \, C_{n-\ell-1}^{\ell+1} 
   \left( \frac {p^{2}-\beta^{2}} {p^{2}+\beta^{2}} \right) \, 
    \mathcal{Y}_{\ell}^{m} (- \mathrm{i} \bm{p}) \, .
\end{align}
To obtain the Fourier transform of a normalized Sturmian defined by
\cref{Def_SturmFun}, we multiply \cref{FT_Sturm_13} by the normalization
factor
$(2 \beta)^{3/2} \, \left[ (n-\ell-1)! / [2n(n+\ell)!] \right]^{1/2}$,
yielding \cite[Eq.\ (4.24)]{Weniger/1985}:
\begin{align}
  \label{FT_SturmFun}
  & \overline{\Psi_{n, \ell}^{m}} (\beta, \bm{p}) \; = \; (2 \pi)^{-3/2}
   \, \int \, \mathrm{e}^{\mathrm{i} \bm{p} \cdot \bm{r}} \, 
    \Psi_{n, \ell}^{m} (\beta, \bm{r}) \, \mathrm{d}^{3} \bm{r}
  \notag
  \\
  & \quad \; = \; 2^{\ell} \ell! \, \, \left[ \frac 
   {2\beta n (n-\ell-1)!}{\pi (n+\ell)!} \right]^{1/2} \left[ \frac 
    {2\beta} {p^{2}+\beta^{2}} \right]^{\ell+2}
  \notag
  \\
  & \qquad \times 
   C_{n-\ell-1}^{\ell+1} \left( \frac {p^{2}-\beta^{2}}
    {p^{2}+\beta^{2}} \right) \, 
     \mathcal{Y}_{\ell}^{m} (- \mathrm{i} \bm{p}) \, .
\end{align}
To obtain the Fourier transform of a bound-state hydrogen eigenfunction,
we only have to use \cref{SturmFun<->BSHEF} and make the substitution
$\beta \mapsto Z/n$. Thus, we obtain \citep*[Eq.\
(28)]{Podolsky/Pauling/1929}:
\begin{align}
  \label{FT_HydEigFun}
  & \overline{W_{n, \ell}^{m}} (Z, \bm{p}) \; = \; (2 \pi)^{-3/2} \, 
   \int \, \mathrm{e}^{\mathrm{i} \bm{p} \cdot \bm{r}} \, 
    W_{n, \ell}^{m} (Z, \bm{r}) \, \mathrm{d}^{3} \bm{r}
  \notag
  \\
  & \quad \; = \; 2^{\ell} \ell! \, \, \left[ \frac 
   {2Z (n-\ell-1)!}{\pi (n+\ell)!} \right]^{1/2} \left[ \frac 
    {2 Z n} {n^{2} p^{2}+Z^{2}} \right]^{\ell+2}
  \notag
  \\ 
  & \qquad \times  
    C_{n-\ell-1}^{\ell+1} 
     \left( \frac {n^{2} p^{2}-Z^{2}} {n^{2} p^{2}+Z^{2}} \right) \, 
      \mathcal{Y}_{\ell}^{m} (- \mathrm{i} \bm{p}) \, .  
\end{align}
This Fourier transformation was in principle also derived by \citet[Eq.\
(26) on p.\ 241]{Rotenberg/1970} in disguised form. However,
\citeauthor{Rotenberg/1970}'s results are misleading because of an
unfortunate definition of the Sturmians (compare \citep[p.\
283]{Weniger/1985}).

If we compare \cref{FT_HydEigFun} with formulas published by other
authors, we find some discrepancies. In the formula given by \citet[Eq.\
(28)]{Podolsky/Pauling/1929}, a phase factor $(- \mathrm{i})^{\ell}$ is
missing. The same error was reproduced by \citet[Eq.\
(8.8)]{Bethe/Salpeter/1977}. The formula given by \citet[Eqs.\ (5.5) and
(5.6)]{Englefield/1972} differs from \cref{FT_HydEigFun} by a phase
factor $(-1)^{m}$. Finally, in the expression given by \citet*[Eq.\
(7.4.69)]{Biedenharn/Louck/1981a} a factor $\pi^{-1/2}$ is missing.

\citet*[pp.\ 50 - 52]{Kaijser/Smith/1977} showed that the generating
function approach of \citeauthor{Podolsky/Pauling/1929} can be extended
to the Fourier transform of a Lambda function defined by
\cref{Def_LambdaFun}. However, the approach of
\citet*{Podolsky/Pauling/1929} and \citet*{Kaijser/Smith/1977} requires
considerable manipulative skills. In \cref{Sec:ExpRBF}, I will show how
the Fourier transforms of bound-state hydrogen eigenfunctions, Sturmians,
and Lambda functions and of other Laguerre-type functions can be
constructed in an almost trivially simple way by expanding generalized
Laguerre polynomials in terms of so-called reduced Bessel functions
(compare \cite[Section IV]{Weniger/1985}).

%
\typeout{==> Section: The Work of Y\"{u}k\c{c}\"{u} and
  Y\"{u}k\c{c}\"{u}}
\section{The Work of Y\"{u}k\c{c}\"{u} and Y\"{u}k\c{c}\"{u}}
\label{Sec:WorkYuekcueYuekcue}
%

\citet*{Podolsky/Pauling/1929} faced the problem that no simple closed
form expression for the Hankel-type integral in \cref{FT_Sturm_1} was
known. They solved this problem by computing instead the Fourier
transform of the generating function \eqref{GLag_GenFun}, which leads to
the comparatively simple and explicitly known Hankel-type integral in
\cref{FT_Sturm_2}. In this way, \citeauthor{Podolsky/Pauling/1929} only
had to perform a series expansion of the radial integral in
\cref{FT_Sturm_2} to derive the explicit expression \eqref{FT_HydEigFun}
for the Fourier transform of a bound state hydrogen eigenfunction.

\citet*{Yuekcue/Yuekcue/2018}, who were apparently unaware of the work by
\citet*{Podolsky/Pauling/1929} or of the whole extensive literature on
this topic, proceeded differently. They utilized the fact that a
generalized Laguerre polynomial $L_{n}^{(\alpha)} (z)$ is according to
\cref{GLag_1F1} a polynomial of degree $n$ in $z$. Thus, the generalized
Laguerre polynomial $L_{n-\ell-1}^{(2\ell+1)} (2 \beta r)$ occurring in
\cref{Def_SturmFun} can be expressed as a sum of powers:
\begin{align}
  \label{GLag->PowZ}
  & L_{n-\ell-1}^{(2\ell+1)} (2 \beta r) 
  \notag
  \\
  & \quad \; = \; 
   \frac {(n+\ell)!} {(n-\ell-1)!} \, \sum_{\nu=0}^{n-\ell-1} \,
    \frac {(-n+\ell+1)_{\nu}}{(2\ell+\nu+1)!} \, 
     \frac {(2 \beta r)^{\nu}}{\nu!} \, .
\end{align}

To achieve what they believe to be a further simplification,
\citet*[Eqs.\ (10) - (12)]{Yuekcue/Yuekcue/2018} combined
\cref{GLag->PowZ} with the Laguerre multiplication theorem \citep*[p.\
249]{Magnus/Oberhettinger/Soni/1966}
\begin{equation}
  \label{GLag_MultThm}
  L_{n}^{(\alpha)} (z x) \; = \; \sum_{m=0}^{n} \, 
   \binom{n+\alpha}{m} \, \frac{(1-z)^{m}}{z^{m-n}} \,
    L_{n-m}^{(\alpha)} (x) \, .
\end{equation}
However, the combination of \cref{GLag->PowZ,GLag_MultThm} leads to the
same Hankel-type integrals as the direct use of \cref{GLag->PowZ}. Thus,
this combination accomplishes nothing and only introduces a completely
useless additional inner sum. Therefore, I will only consider the direct
use of \cref{GLag->PowZ}.

In \citeyear{Slater/1930}, \citet{Slater/1930} introduced the so-called
Slater-type functions, which had an enormous impact on atomic electronic
structure theory and which in unnormalized form are expressed as follows:
\begin{align}
  \label{Def_STF}
  \chi_{n, \ell}^{m} (a, \bm{r}) \; = \; & (\alpha r)^{n-1} \,
   \mathrm{e}^{-\alpha r} \, Y_{\ell}^{m} (\theta, \phi)
  \notag
  \\
  \; = \; & 
   (\alpha r)^{n-\ell-1} \, \mathrm{e}^{-\alpha r} \, 
    \mathcal{Y}_{\ell}^{m} (\alpha \bm{r}) \, ,
     \quad \alpha > 0 \, .
\end{align}
I always tacitly assume that the principal quantum number $n$ is a
positive integer $n \in \mathbb{N}$ satisfying $n - \ell \ge 1$.

With the help of Slater-type functions, an unnormalized Sturmian function
\eqref{Def_SturmFun} with fixed $\beta > 0$ can be expressed as follows:
\begin{align}
  \label{SturmFunUN->STF}
  & \mathrm{e}^{-\beta r}  \, L_{n-\ell-1}^{(2\ell+1)} (2\beta r) \,
     \mathcal{Y}_{\ell}^{m} (2 \beta \bm{r}) 
      \; = \; \frac {(n+\ell)!} {(n-\ell-1)!}
  \notag
  \\
  & \quad \, \times \, \sum_{\nu=0}^{n-\ell-1} \,
    \frac {(-n+\ell+1)_{\nu}}{(2\ell+\nu+1)!} \, 
     \frac {2^{\nu}}{\nu!} \, 
      \chi_{\nu+\ell+1, \ell}^{m} (\beta, \bm{r}) \, .
\end{align}

The idea of expressing functions based on the generalized Laguerre
polynomial by finite sums of Slater-type functions is not new. To the
best of my knowledge, it was introduced by \citet{Smeyers/1966} in
\citeyear{Smeyers/1966}, who expressed Lambda functions defined by
\cref{Def_LambdaFun} as linear combinations of Slater-type
functions. \citeauthor{Smeyers/1966} constructed in this way one-range
addition theorems of Slater-type functions, which were expansion in terms
of Lambda functions. Thus, their expansion coefficients are overlap
integrals \citep[Section 3]{Smeyers/1966}. In \citeyear{Guseinov/1978},
\citet[Eqs.\ (6) - (8)]{Guseinov/1978} adopted \citeauthor{Smeyers/1966}'
approach and consistently used it in his countless later publications,
without ever giving credit to \citet{Smeyers/1966}.

Smeyers' approach is undoubtedly very simple. Nevertheless, it is not
good. In \cref{GLag->PowZ} there are strictly alternating
signs. Therefore, in sums of the type of \cref{SturmFunUN->STF}, which
inherit the alternating signs from \cref{GLag->PowZ}, numerical
instabilities are to be expected in the case of larger summation
indices. This had already been emphasized in
\citeyear{Trivedi/Steinborn/1982} by \citet*[pp.\ 116 -
117]{Trivedi/Steinborn/1982}. For a more detailed discussion plus
additional references, see \citep[pp.\ 32 - 34]{Weniger/2012}.

Fourier transformation is a linear operation. Consequently,
\cref{SturmFunUN->STF} implies that the Fourier transformation of a
Sturmian -- or of any of the various other functions based on generalized
Laguerre polynomials -- can be expressed as a finite linear combination
of Fourier transforms of Slater-type functions with integral principal
quantum numbers (compare \citep*[Eq.\ (12)]{Yuekcue/Yuekcue/2018}).

There is an extensive literature on Fourier transforms of Slater-type
functions. I am aware of articles by \citet{Geller/1962,Geller/1963a},
\citet{Silverstone/1966,Silverstone/1967a},
\citet*{Edwards/Gottlieb/Doddrell/1979}, \citet*{Henneker/Cade/1968},
\citet*{Kaijser/Smith/1977}. \citet*{Weniger/Steinborn/1983a},
\citet{Niukkanen/1984c}, \citet*{Belkic/Taylor/1989}, and by
\citet{Akdemir/2018}. In addition, there is a Wikipedia article
\citep{WikipediaSTF/2018}, whose principal reference is the article by
\citet*{Belkic/Taylor/1989}. \citet*{Yuekcue/Yuekcue/2018} only mentioned
\citet{Niukkanen/1984c} as their Ref.\ [8].

The Rayleigh expansion \eqref{Rayleigh_Expan} leads to an expression of
the Fourier transform of a Slater-type function as a Hankel-type radial
integral:
\begin{align}
  \label{Def_STF_FT}
  & \overline{\chi_{n,\ell}^{m}} (\alpha, \bm{p}) \; = \; (2\pi)^{-3/2} 
   \, \int \, \mathrm{e}^{- \mathrm{i} \bm{p} \cdot \bm{r}} \, 
    \chi_{n,\ell}^{m} (\alpha, \bm{r}) \, \mathrm{d}^3 \bm{r}  
  \notag
  \\
  & \qquad \; = \;
  \alpha^{n-1} \, (-\mathrm{i})^{\ell} \, Y_{\ell}^{m} (\bm{p}/p) 
  \notag 
  \\
  & \qquad \quad \; \; \times \, 
   p^{-1/2} \int_{0}^{\infty} \, \mathrm{e}^{- \alpha r} \, 
    r^{n+1/2} \, J_{\ell+1/2} (p r) \, \mathrm{d} r \, . 
\end{align}
This Hankel-type integral is a special case of the one in
\cref{Int_ExpBesJ}. Thus, we immediately obtain \citep*[Eq.\
(3.15)]{Weniger/Steinborn/1983a}
\begin{align}
  \label{FT_STF_1}
  & \overline{\chi_{n,\ell}^{m}} (\alpha, \bm{p}) \; = \; \frac 
   {(n+\ell+1)!} {\alpha^{\ell+3} \, (2\pi)^{1/2} \, (1/2)_{\ell+1}} \,
    \mathcal{Y}_{\ell}^{m} (-\mathrm{i} \bm{p}/2) 
  \notag 
  \\
  & \qquad \times \, {}_{2} F_{1} \left( \frac{n+\ell+2}{2},
   \frac{n+\ell+3}{2}; \ell+\frac{3}{2}; - \frac{p^2}{\alpha^2} \right)
    \, ,
\end{align}
which corresponds to \citep*[Eqs.\ (14) and (16)]{Yuekcue/Yuekcue/2018}.

For the derivation of Watson's hypergeometric representation
\eqref{Int_ExpBesJ}, one only has to insert the power series
$J_{\nu} (z) = \bigl[ (z/2)^{\nu}/\Gamma(\nu+1) \bigr] {}_{0} F_{1}
(\nu+1; -z^{2}/4)$ \citep*[Eq.\
(10.16.9)]{Olver/Lozier/Boisvert/Clark/2010} into the integral, followed
by an interchange of summation and term-wise integration. In the final
step, the Pochhammer duplication formula \citep*[Eq.\
(5.2.8)]{Olver/Lozier/Boisvert/Clark/2010} is to be used.

\citeauthor{Watson/1922}'s derivation can be modified easily. If we
instead use
$J_{\nu} (z) = \bigl[ (z/2)^{\nu} \mathrm{e}^{\mp \mathrm{i} z}/\Gamma
(\nu+1) \bigr] {}_{1} F_{1} (\nu+1/2; 2\nu+1; \pm 2 \mathrm{i} z)$
\citep*[Eq.\ (10.16.5)]{Olver/Lozier/Boisvert/Clark/2010}, we obtain an
alternative representation which involves a ${}_{2} F_{1}$ with complex
argument:
\begin{align}
  \label{Int_ExpBesJ_CompArg}
  & \int_{0}^{\infty} \, \mathrm{e}^{-at} \, J_{\nu} (bt) \, t^{\mu-1}
   \, \mathrm{d} t \; = \;  \frac 
   {(b/2)^{\nu}}{(a \pm \mathrm{i} b)^{\mu+\nu}} \, 
    \frac{\Gamma (\mu+\nu)}{\Gamma (\nu+1)}
  \notag
  \\
  &  \quad \times \, {}_{2} F_{1}
     \left( \nu+\frac{1}{2}, \mu+\nu; 2\nu+1; 
      \frac{\pm 2 \mathrm{i} b}{a \pm \mathrm{i} b} \right) \, . 
\end{align}
This yields the following Fourier transformation:
\begin{align}
  \label{FT_STF_1_C}
  & \overline{\chi_{n,\ell}^{m}} (\alpha, \bm{p}) \; = \; 
   \frac {(n+\ell+1)!} {(2\pi)^{1/2} \, (1/2)_{\ell+1}} \, 
    \frac{\alpha^{n-1}}{(\alpha \pm \mathrm{i} p)^{n+\ell+2}} \, \mathcal{Y}_{\ell}^{m} (-\mathrm{i} \bm{p}/2)
  \notag
  \\ 
  & \qquad \times \, 
   {}_{2} F_{1} \left( n+\ell+2, \ell+1; 2 \ell +2; 
     \frac{\pm 2 \mathrm{i} p}{\alpha \pm \mathrm{i} p} \right) \, . 
\end{align}
A slightly less general result had been obtained by \citet[Eq.\
(15)]{Belkic/Taylor/1989}. They used only the upper signs in the
expression involving the ${}_{1} F_{1}$ given above. In this way,
\citet[Eq.\ (15)]{Belkic/Taylor/1989} obtained only the upper signs on
the right-hand side of \cref{FT_STF_1_C}.

The Hankel-type integral in \cref{Int_ExpBesJ_CompArg} is real if
$\mu, \nu, a, p \in \mathbb{R}$. Thus, the right-hand side of
\cref{Int_ExpBesJ_CompArg} also has to be real, or equivalently, it has
to be equal to its complex conjugate. This implies:
\begin{align}
  \label{HTI_mod_4_real}
  & {}_{2} F_{1}
    \left( \nu+\frac{1}{2}, \mu+\nu; 2\nu+1; 
     \frac{\pm 2 \mathrm{i} b}{a \pm \mathrm{i} b} \right) \; = \; 
      \left( \frac{a \pm \mathrm{i} b} 
       {a \mp \mathrm{i} b} \right)^{\mu+\nu}
  \notag
  \\
  & \quad \times \, 
    {}_{2} F_{1} \left( \nu+\frac{1}{2}, \mu+\nu; 2\nu+1; 
      \frac{\mp 2 \mathrm{i} b}{a \mp \mathrm{i} b} \right) \, . 
\end{align}
Analogous symmetries also exist in \cref{FT_STF_1_C} and in all other
expressions of that kind derived later.

The hypergeometric series ${}_{2} F_{1}$ in \cref{FT_STF_1,FT_STF_1_C} do
not converge for all $p \ge 0$. We either have
$\vert-p^{2}/\alpha^{2} \vert \to \infty$ as $p \to \infty$, or
$\vert \pm 2 \mathrm{i} p/(\alpha \pm \mathrm{i} p) \vert \to 2$ as
$p \to \infty$. Thus, \cref{FT_STF_1,FT_STF_1_C} are not sufficient for
computational purposes. They are, however, convenient starting points for
the construction of alternative expressions with better numerical
properties. In the case of the ${}_{2} F_{1}$ in \cref{FT_STF_1}, this
had already been emphasized in the first edition of
\citeauthor{Watson/1922}'s classic book \citep[Eq.\ (3) on p.\
385]{Watson/1922} which appeared in \citeyear{Watson/1922}.

For the construction of analytic continuation formulas, it makes sense to
use the highly developed transformation theory of the Gaussian
hypergeometric function ${}_{2} F_{1}$ as an ordering principle (see for
example \citep*[pp.\ 47 - 51]{Magnus/Oberhettinger/Soni/1966} or
\citep*[\S 15.8 Transformations of
Variable]{Olver/Lozier/Boisvert/Clark/2010}). This leads to a vast number
of alternatively expressions (far too many to be presented
here). Therefore, I will only concentrate on illustrative examples.

The first author -- N.\ \citeauthor{Yuekcue/2017} -- should be aware of
the relevance of analytic continuation formulas of hypergeometric
functions because of his recent article \emph{Hypergeometric Functions in
  Mathematics and Theoretical Physics} \citep{Yuekcue/2017}.

The simplest transformations of a ${}_{2} F_{1}$ are the Euler and Pfaff
transformations (see for example \citep*[p.\
47]{Magnus/Oberhettinger/Soni/1966}):
\begin{align}
  \label{LTr_0}
  & {}_{2} F_{1} (a, b; c; z)
  \notag
  \\
  & \qquad \; = \;
   (1-z)^{c-a-b} \, {}_{2} F_{1} (c-a, c-b; c; z)
  \\
  \label{LTr_1}
  & \qquad \; = \;
   (1-z)^{-a} \, {}_{2} F_{1} \bigl( a, c-b; c; z/(z-1) \bigr)
  \\
  \label{LTr_2}
  & \qquad \; = \;
   (1-z)^{-b} \, {}_{2} F_{1} \bigl( c-a, b; c; z/(z-1) \bigr) \, .
\end{align}

The application the Euler transformation \eqref{LTr_0} to the
${}_{2} F_{1}$s in \cref{FT_STF_1,FT_STF_1_C} yields \citep*[Eq.\
(3.11)]{Weniger/Steinborn/1983a}
\begin{align}
  \label{FT_STF_2}
  & \overline{\chi_{n,\ell}^{m}} (\alpha, \bm{p}) \; = \; \frac 
   {(n+\ell+1)!} {(2\pi)^{1/2} \, (1/2)_{\ell+1}} \,
    \mathcal{Y}_{\ell}^{m} (-\mathrm{i} \bm{p}/2) 
  \notag 
  \\
  & \quad \times \, \frac{\alpha^{2n-\ell-1}}{[\alpha^2+p^2]^{n+1}} 
  \notag 
  \\
  & \qquad \times \, \,
   {}_{2} F_{1} \left( \frac{\ell-n}{2}, \frac{\ell-n+1}{2}; 
    \ell+\frac{3}{2}; - \frac{p^2}{\alpha^2} \right)
\end{align}
and
\begin{align}
\label{FT_STF_1_C_Euler}
  & \overline{\chi_{n,\ell}^{m}} (\alpha, \bm{p}) \; = \; \frac
   {(n+\ell+1)!} {(2\pi)^{1/2} \, (1/2)_{\ell+1}} \,  
    \mathcal{Y}_{\ell}^{m} (-\mathrm{i} \bm{p}/2)
  \notag
  \\
  & \qquad \times \, \,
   \frac{\alpha^{n-1}} {(\alpha \pm \mathrm{i} p)^{\ell+1} \, 
    (\alpha \mp \mathrm{i} p)^{n+1}} 
  \notag 
  \\
   & \qquad \quad \, {}_{2} F_{1} \left( \ell-n, \ell+1; 2 \ell +2; 
      \frac{\pm 2 \mathrm{i} p}{\alpha \pm \mathrm{i} p} \right) \, .
\end{align}
Since we assume $n \in \mathbb{N}$, $\ell \in \mathbb{N}_{0}$,
$n-\ell-1 \ge 0$, $n-\ell$ and either $(\ell-n)/2$ or $(\ell-n+1)/2$ are
positive integers. Accordingly, the ${}_{2} F_{1}$s in
\cref{FT_STF_2,FT_STF_1_C_Euler} terminate, which represents a
substantial improvement compared to the non-terminating ${}_{2} F_{1}$s
in \cref{FT_STF_1,FT_STF_1_C}. Both \cref{FT_STF_2,FT_STF_1_C_Euler}
allow a convenient evaluation of
$\overline{\chi_{n,\ell}^{m}} (\alpha, \bm{p})$ for all
$\bm{p} \in \mathbb{R}^{3}$. For recurrence formulas of the Gaussian
hypergeometric function ${}_{2} F_{1} (a, b; c;z)$, where two or three of
the parameters $a$, $b$, and $c$ change \emph{simultaneously}, see
\citep[Appendix C]{Weniger/2001}.

We can also employ the Pfaff transformations \eqref{LTr_1} and
\eqref{LTr_2}. In the case of \cref{FT_STF_1}, this yields hypergeometric
series with argument $p^{2}/(\alpha^{2}+p^{2})$ that either terminate or
converge for all $p \ge 0$ \citep*[Eqs.\ (3.16) and
(3.17)]{Weniger/Steinborn/1983a}. In the case of \cref{FT_STF_1_C_Euler},
we only obtain complex conjugates of known radial parts.

But this is not yet the end of the story. By systematically exploiting
the \emph{known} transformation properties of the Gaussian hypergeometric
function ${}_{2} F_{1}$, many other terminating or non-terminating
expressions can be derived. For example, we could also use one of the
linear transformations that accomplish the variable transformations
$z \mapsto 1-z$, $z \mapsto 1/z$, $z \mapsto 1/(1-z)$, and
$z \mapsto 1 - 1/z$, respectively, by expressing a given ${}_{2} F_{1}$
in terms of two other ${}_{2} F_{1}$s (see for example \citep*[pp.\ 47 -
49]{Magnus/Oberhettinger/Soni/1966}). Normally, these transformations
lead to comparatively complicated expressions which can safely be
ignored. An exception is the following expression obtained by a
transformation $z \mapsto 1/(1-z)$ \citep*[Eqs.\ (3.19) and
(3.20)]{Weniger/Steinborn/1983a}:
{\allowdisplaybreaks
\begin{align}
  \label{FT_STF_5}
  & \overline{\chi_{n,\ell}^{m}} (\alpha, \bm{p}) \; = \; (\pi/2)^{1/2} 
   \, (n+\ell+1)! \, \mathcal{Y}_{\ell}^{m} (-\mathrm{i} \bm{p}/2)
  \notag 
  \\
  & \quad \times \, \Biggl[
  \frac{\alpha^{n-1} [\alpha^2+p^2]^{-(n+\ell+2)/2}}{\Gamma
   \bigl([n+\ell+3]/2 \bigr) \Gamma \bigl([\ell-n+1]/2 \bigl)}  
  \notag 
  \\[1\jot]
  & \qquad \times \, {}_{2} F_{1} \left( \frac{n+\ell+3}{2},
    \frac{\ell-n}{2}; \frac{1}{2}; \frac{\alpha^2}{\alpha^2+p^2} \right)
  \notag \\[1\jot]
  & \quad - \, \frac{2\alpha^{n} [\alpha^2+p^2]^{-(n+\ell+3)/2}}
   {\Gamma \bigl([n+\ell+2]/2 \bigr) \Gamma \bigl([\ell-n]/2 \bigr)}
  \notag \\[1\jot]
  & \qquad \times \, {}_{2} F_{1} \left( \frac{n+\ell+3}{2},
    \frac{\ell-n+1}{2}; \frac{3}{2}; \frac{\alpha^2}{\alpha^2+p^2}
  \right) \Biggr] \, .
\end{align} }
This expression is simpler than it looks. If $n-\ell$ is even, the second
part of the right-hand side vanishes because of the gamma function
$\Gamma \bigl([\ell-n]/2 \bigr)$, and if $n-\ell$ is odd, the first part
vanishes because of the gamma function
$\Gamma \bigl([\ell-n+1]/2 \bigr)$.

In addition to linear transformations, a Gaussian hypergeometric function
${}_{2} F_{1}$ may also satisfy so-called quadratic transformations (see
for example \citep*[pp.\ 49 - 51]{Magnus/Oberhettinger/Soni/1966} or
\citep*[\S 15.8(iii) Quadratic
Transformations]{Olver/Lozier/Boisvert/Clark/2010}). Unlike the linear
transformations considered so far, quadratic transformations do not exist
for a ${}_{2} F_{1} (a, b; c; z)$ with completely arbitrary parameters
$a$, $b$, and $c$. They only exists for special values of the parameters
$a$, $b$, and $c$ \citep*[Table
15.8.1]{Olver/Lozier/Boisvert/Clark/2010}.

The hypergeometric series in \cref{FT_STF_1} is of the general type
${}_{2} F_{1} \left( a, a+1/2; c; z \right)$. This suggests the
application of the following quadratic transformations \citep*[p.\
50]{Magnus/Oberhettinger/Soni/1966} to this ${}_{2} F_{1}$:
\begin{align}
  \label{QT_2F1_17}
  & {}_{2} F_{1} \left( a, a+\frac{1}{2}; c; z \right) 
  \notag
  \\
  & \quad \; = \; (1-z)^{-a} \, {}_{2} F_{1} 
   \left( 2a, 2c-2a-1; c; \frac{\sqrt{1-z}-1}{2\sqrt{1-z}} \right)
  \\
  \label{QT_2F1_18}
  & \quad \; = \; \left(1\pm\sqrt{z}\right)^{-2a} \, 
   {}_{2} F_{1} \left( 2a, c-\frac{1}{2}; 2c-1;  
    \pm \frac{2\sqrt{z}}{1 \pm \sqrt{z}} \right)
  \\
  \label{QT_2F1_19}
  & \quad \; = \; \left( \frac{1+\sqrt{1-z}}{2} \right)^{-2a}
  \notag
  \\  
  & \qquad \quad \times \, {}_{2} F_{1} \left( 2a, 2a-c+1; c;  
   \frac{1-\sqrt{1-z}}{1+\sqrt{1-z}} \right)
\end{align}
Application of \cref{QT_2F1_17,QT_2F1_18,QT_2F1_19} to the ${}_{2} F_{1}$
in \cref{FT_STF_1} yields the following alternative expressions:
\begin{align}
  \label{FT_STF_1_QT_17}
  & \overline{\chi_{n,\ell}^{m}} (\alpha, \bm{p})
  \notag 
  \\
  & \quad \; = \; \frac 
   {(n+\ell+1)!} {(2\pi)^{1/2} \, (1/2)_{\ell+1}} \, 
    \frac{\alpha^{n-1}}{[\alpha^{2} + p^{2}]^{(n+\ell+2)/2}} \,
     \mathcal{Y}_{\ell}^{m} (-\mathrm{i} \bm{p}/2)  
  \notag
  \\
  & \qquad \times \, {}_{2} F_{1} \left( n+\ell+2, \ell-n;
   \ell+\frac{3}{2}; \frac{\sqrt{\alpha^{2}+p^{2}}-\alpha}     {2\sqrt{\alpha^{2}+p^{2}}} \right)
  \\
  \label{FT_STF_1_QT_18}
  & \quad \; = \; \frac 
   {(n+\ell+1)!} {(2\pi)^{1/2} \, (1/2)_{\ell+1}} \, 
    \frac{\alpha^{n-1}}{(\alpha \pm \mathrm{i} p)^{n+\ell+2}} \,
     \mathcal{Y}_{\ell}^{m} (-\mathrm{i} \bm{p}/2)  
  \notag
  \\
  & \qquad \times \, 
   {}_{2} F_{1} \left( n+\ell+2, \ell+1; 2\ell+2;
    \frac{ \pm 2\mathrm{i}p}{\alpha \pm \mathrm{i} p} \right)
  \\
  \label{FT_STF_1_QT_19}
  & \quad \; = \; \frac {(n+\ell+1)!} {(2\pi)^{1/2} \, (1/2)_{\ell+1}} \,
    \frac {2^{n+\ell+2} \, \alpha^{n-1}} 
     {[\alpha^{2}+p^{2}]^{(n+\ell+2)/2}} \,
      \mathcal{Y}_{\ell}^{m} (-\mathrm{i} \bm{p}/2)  
  \notag
  \\
  & \qquad \times \,
    {}_{2} F_{1} \left( n+\ell+2, n+\frac{3}{2}; \ell+\frac{3}{2};  
     \frac{\alpha-\sqrt{\alpha^{2}+p^{2}}}
      {\alpha+\sqrt{\alpha^{2}+p^{2}}} \right) \, . 
\end{align}
\Cref{FT_STF_1_C,FT_STF_1_QT_18} are identical. The derivation of
\cref{FT_STF_1_QT_18} shows that the quadratic transformation
\eqref{QT_2F1_18} can create a representation containing a ${}_{2} F_{1}$
with complex argument from a ${}_{2} F_{1}$ with real
argument. Consequently, the derivation of the complex expression
\eqref{Int_ExpBesJ_CompArg} for the Hankel-type integral in
\cref{Int_ExpBesJ} by directly evaluating the integral is -- strictly
speaking -- superfluous. Applying the quadratic transformation
\eqref{QT_2F1_18} to the ${}_{2} F_{1}$ in \cref{Int_ExpBesJ} would have
done the job.

\begin{widetext}
  Only the ${}_{2} F_{1}$ in \cref{FT_STF_1_QT_17} terminates. As a
  remedy, we can apply the Euler transformation \eqref{LTr_0} to the
  non-terminating ${}_{2} F_{1}$s in
  \cref{FT_STF_1_QT_18,FT_STF_1_QT_19}, yielding
  \begin{align}
    \label{FT_STF_1_QT_18_Euler}
    & \overline{\chi_{n,\ell}^{m}} (\alpha, \bm{p})
    \notag
    \\
    & \qquad \; = \;  
      \frac {(n+\ell+1)!} {(2\pi)^{1/2} \, (1/2)_{\ell+1}} \,
       \frac{\alpha^{n-1}} {(\alpha \pm \mathrm{i} p)^{\ell+1} \, 
        (\alpha \mp \mathrm{i} p)^{n+1}} \, 
         \mathcal{Y}_{\ell}^{m} (-\mathrm{i} \bm{p}/2) \,
          {}_{2} F_{1} \left( \ell-n, \ell+1; 2 \ell +2; 
           \frac{\pm 2 \mathrm{i} p}{\alpha \pm \mathrm{i} p} \right)
     \\ 
    \label{FT_STF_1_QT_19_Euler}
    & \qquad \; = \; \frac {(n+\ell+1)!}{(2\pi)^{1/2} \, (1/2)_{\ell+1}}
     \, \frac{\alpha^{n-1}}{2^{n-\ell}} \,
      \frac{\left[\alpha+\sqrt{\alpha^{2}+p^{2}} \right]^{2n+2}}
       {\left[\alpha^{2}+p^{2}\right]^{(3n+\ell+4)/2}} \, 
        \mathcal{Y}_{\ell}^{m} (-\mathrm{i} \bm{p}/2) \,
         {}_{2} F_{1} \left(\ell-n, -n-\frac{1}{2}; \ell+\frac{3}{2};  
          \frac{\alpha-\sqrt{\alpha^{2}+p^{2}}}
           {\alpha+\sqrt{\alpha^{2}+p^{2}}} \right) \, .  
  \end{align}  
\end{widetext}

The terminating ${}_{2} F_{1}$ in \cref{FT_STF_1_QT_17} can be expressed
as a Gegenbauer polynomial via \citep*[p.\
220]{Magnus/Oberhettinger/Soni/1966}
\begin{equation}
  \label{GegPol_2F1_a}
  C_{n}^{\lambda} (x) \; = \; \frac{(2\lambda)_n}{n!} \, 
   {}_2 F_1 \left( -n, n+2\lambda; \lambda+\frac{1}{2}; 
    \frac{1-x}{2} \right) \, ,
\end{equation}
yielding
\begin{align}
  \label{FT_STF_8_GegPol}
  & \overline{\chi_{n,\ell}^{m}} (\alpha, \bm{p}) \; = \; 
   \frac {(n+\ell+1)! \, (n-\ell)!} 
    {(2\pi)^{1/2} \, (1/2)_{\ell+1} \, (2\ell+2)_{n-\ell}} \, 
     \mathcal{Y}_{\ell}^{m} (-\mathrm{i} \bm{p}/2)
  \notag
  \\
  & \qquad \times \,
   \frac{\alpha^{n-1}}{[\alpha^2+p^2]^{\frac{n+\ell}{2}+1}} \,
    C_{n-\ell}^{\ell+1}  
     \left( \frac{\alpha}{\sqrt{\alpha^{2}+p^{2}}} \right) \, .
\end{align}
The Gegenbauer polynomial representation \eqref{FT_STF_8_GegPol}
corresponds to the second representation given by
\citet*{Yuekcue/Yuekcue/2018} in their Eqs.\ (13) and (15). As their
source, \citet*{Yuekcue/Yuekcue/2018} give the book by
\citet*{Gradshteyn/Rhyzhik/2000} as their Ref.\ [25], without specifying
a page or equation number. Unfortunately, I was not able to find the
corresponding expression in the book by \citet*{Gradshteyn/Rhyzhik/2000}.

In earlier articles by \citet*[Eq.\
(12)]{Yavuz/Yuekcue/Oeztekin/Yilmaz/Doenduer/2005} and \citet[Eq.\
(32)]{Yuekcue/2015}, the Gegenbauer polynomial representation
\eqref{FT_STF_8_GegPol} had been attributed to
\citet{Guseinov/1987}. Google Scholar gave me the title of
\citeauthor{Guseinov/1987}'s article, but I was not able to obtain a
copy. Personal contacts to the Wuhan Institute of Physics and Mathematics
of the Chinese Academy of Sciences could not help, either.

But \citeauthor{Guseinov/1987} was not the first one to derive the
Gegenbauer polynomial representation \eqref{FT_STF_8_GegPol}. To the best
of my knowledge, this had been achieved by \citet{Niukkanen/1984c} in
\citeyear{Niukkanen/1984c}, who introduced a fairly large class of
exponentially decaying functions \citep[Eqs.\ (2) and
(3)]{Niukkanen/1984c}, which contain all function sets considered in this
article as special cases. The radial part of the Fourier transform of
\citeauthor{Niukkanen/1984c}'s function can be expressed in terms of an
Appell function $F_{2}$ \citep[Eq.\ (21)]{Niukkanen/1984c}, which is an
hypergeometric function in two variables \citep*[Eq.\
(16.13.2)]{Olver/Lozier/Boisvert/Clark/2010}. By means of a reduction
formula in combination with a suitable quadratic transformation of a
${}_{2} F_{1}$, \citet[Eq.\ (55)]{Niukkanen/1984c} obtained the
Gegenbauer polynomial representation \eqref{FT_STF_8_GegPol}. This
Gegenbauer representation had also been derived by \citet*[Eq.\
(21)]{Belkic/Taylor/1989} in \citeyear{Belkic/Taylor/1989} in connection
with their restricted version of \cref{Int_ExpBesJ_CompArg} \citep[Eq.\
(15)]{Belkic/Taylor/1989}.

\citet*{Yuekcue/Yuekcue/2018} used either a representation given by their
Eqs.\ (14) and (16) involving a non-terminating ${}_{2} F_{1}$, which
correspond to \cref{FT_STF_1}, or alternatively a Gegenbauer polynomial
representation given by their Eqs.\ (13) and (15), which correspond to
\cref{FT_STF_8_GegPol}. The non-terminating ${}_{2} F_{1}$ in
\cref{FT_STF_1} converges only for $\vert p^{2}/\alpha^{2} \vert < 1$,
whereas the Gegenbauer polynomial in \cref{FT_STF_8_GegPol} is meaningful
for all $\vert \bm{p} \vert \in \mathbb{R}^{3}$. Thus,
\citeauthor{Yuekcue/Yuekcue/2018} had to prove that their Gegenbauer
polynomial representation provides an analytic continuation of their
representation involving a non-terminating ${}_{2} F_{1}$ with a finite
radius of convergence to all $\vert \bm{p} \vert \in
\mathbb{R}^{3}$. They did this by showing in \citep[Table
1]{Yuekcue/Yuekcue/2018} that the radial parts of these representation
give for a variety of quantum numbers $n$ and $\ell$ and for certain
values of $p$ identical \emph{numerical} results. This highly pedestrian
approach is no substitute for a rigorous mathematical proof.

So far, I only showed that representations involving a ${}_{2} F_{1}$
with real argument can be obtained from representations involving a
${}_{2} F_{1}$ with complex argument (compare
\cref{FT_STF_1_C,FT_STF_1_C_Euler}). However, the inverse operations are
also possible. For example, the application of the quadratic
transformation \citep*[p.\ 51]{Magnus/Oberhettinger/Soni/1966}
\begin{align}
  \label{QT_2F1_32}
  & {}_{2} F_{1} \left( a, b; 2b; z \right) \; = \;
   \left( 1-z/2 \right)^{{-a}}
  \notag
  \\
  & \qquad \times \, 
   {}_{2} F_{1} \left( \frac{a}{2}, \frac{a+1}{2}; b+\frac{1}{2}; 
    \frac{z^{2}}{[2-z]^{2}} \right)  
\end{align}
to the ${}_{2} F_{1}$s in \cref{FT_STF_1_C,FT_STF_1_C_Euler} yields
\cref{FT_STF_1,FT_STF_2}.

By suitably combining linear and quadratic transformations, many explicit
expressions for the Fourier transform of a Slater-type function can be
derived. However, this is not yet the end of the story. Those Gaussian
hypergeometric functions ${}_{2} F_{1}$, for which a quadratic
transformation exists, can also be expressed in terms of Legendre
functions \citep*[pp.\ 51 - 54]{Magnus/Oberhettinger/Soni/1966}. Since,
however, Legendre functions can be viewed to be nothing but special
hypergeometric series ${}_{2} F_{1}$ \citep*[\S 14.3 Definitions and
Hypergeometric Representations]{Olver/Lozier/Boisvert/Clark/2010}, I will
refrain from considering Legendre function representations
explicitly. This would only lead to a repetition of known hypergeometric
expressions in disguise. Let me just mention that \citet*[Eq.\
(6.621.1)]{Gradshteyn/Rhyzhik/2000} expressed the Hankel-type integral in
\cref{Int_ExpBesJ} also in terms of Legendre functions.

My incomplete list of representations of the Fourier transform of a
Slater-type function should suffice to convince even a skeptical reader
that the highly developed transformation theory of the Gaussian
hypergeometric function ${}_{2} F_{1}$ is extremely useful in this
context. It allows the derivation of a large variety of different
representations, which are all analytic continuations of the basic
expressions \eqref{FT_STF_1} and \eqref{FT_STF_1_C}.

The derivation and classification of the various expressions for the
Fourier transforms of Slater-type functions is certainly an achievement
in its own right. Nevertheless, one should not forget that in the context
of the Fourier transform of a bound-state hydrogen eigenfunction or of
other functions based on the generalized Laguerre polynomials, these
Slater results are essentially irrelevant. The formulas presented in this
Section confirm once more what I had already emphasized in \citep*[p.\
29]{Weniger/2012}: although extremely simple in the coordinate
representation, Slater-type functions are comparatively complicated
object in momentum space. Their Fourier transforms have the same level of
complexity as the Fourier transforms of bound state hydrogen
eigenfunctions (see \cite[Section IV]{Weniger/1985}).

Therefore, it cannot be a good idea to express the Fourier transform of a
bound-state hydrogen eigenfunction as a linear combination of Fourier
transforms of Slater-type functions. Because of strictly alternating
sings, \cref{SturmFunUN->STF} as well as all formulas derived from it
become numerically unstable for large quantum numbers $n$. In addition,
these linear combinations of the Fourier transforms of Slater-type
functions are for large $n$ hopelessly inefficient compared to the
classic result \eqref{FT_HydEigFun} derived by \citet*[Eq.\
(28)]{Podolsky/Pauling/1929}. To the best of my knowledge, nobody has
ever been able to construct \cref{FT_HydEigFun} from a linear combination
of Fourier transforms of Slater-type functions.

If we evaluate the Fourier transform of a bound-state hydrogen
eigenfunction or of related functions via linear combinations of the
Fourier transforms of Slater-type functions, we have to deal with
extensive intrinsic cancellations. I learned the hard way from my work on
convergence acceleration and the summation of divergent series (see for
example \citep*{Weniger/1989,Brezinski/RedivoZaglia/Weniger/2010a} or
\citep*[\S 3.9(v) Levin's and Weniger's
Transformations]{Olver/Lozier/Boisvert/Clark/2010}) that expansions,
which are plagued by substantial intrinsic cancellations, can easily
become numerically problematic. It is always desirable to use only those
expressions for computational purposes, whose cancellations had been done
analytically.

%
\typeout{==> Section: Expansion in Terms of Reduced Bessel Functions}
\section{Expansion in Terms of Reduced Bessel Functions}
\label{Sec:ExpRBF}
%

A singe power $z^{n}$ is obviously simpler than a generalized Laguerre
polynomial $L_{n}^{(\alpha)} (z)$. Therefore, it is tempting to believe
that powers produce simpler Hankel-type integrals than corresponding
generalized Laguerre polynomials. However, simplicity is a very elusive
concept, and the results in \cref{Sec:WorkYuekcueYuekcue} show that this
seemingly obvious assumption is not true.

If we want to evaluate the Fourier transforms of bound-state hydrogen
eigenfunctions or of related functions by expanding the generalized
Laguerre polynomials, we must find alternative expansion functions that
have more convenient properties than powers. The so-called reduced Bessel
functions and their an-isotropic generalization, the so-called $B$
functions produce the desired expansions. Based on previous work by
\citet[Eq.\ (55) on p.\ 15]{Shavitt/1963}, $B$ functions were defined in
\citeyear{Filter/Steinborn/1978b} by \citet*[Eq.\
(2.14)]{Filter/Steinborn/1978b} as follows:
\begin{equation}
  \label{Def:B_Fun}
  B_{n,\ell}^{m} (\beta, \bm{r}) \; = \;
   \frac {\hat{k}_{n-1/2} (\beta r)} {2^{n+\ell} (n+\ell)!} \,
    \mathcal{Y}_{\ell}^{m} (\beta \bm{r}) \, .  
\end{equation}
Here, $\beta > 0$, $n \in \mathbb{Z}$, and $\hat{k}_{n-1/2}$ is a reduced
Bessel function. If $K_{\nu} (z)$ is a modified Bessel function of the
second kind \citep*[Eq.\ (10.27.4)]{Olver/Lozier/Boisvert/Clark/2010},
the reduced Bessel function is defined as follows \cite[Eqs.\ (3.1) and
(3.2)]{Steinborn/Filter/1975c}:
\begin{equation}
   \label{Def:RBF}
   \hat{k}_{\nu} (z) \; = \; (2/\pi)^{1/2} \, z^{\nu} \, K_{\nu} (z) \, ,
    \qquad \nu, z \in \mathbb{C} \, .
\end{equation}
If the order $\nu$ is half-integral, $\nu = n + 1/2$ with
$n \in \mathbb{N}_0$, the reduced Bessel function can be expressed as an
exponential multiplied by a terminating confluent hypergeometric series
${}_1 F_1$ (see for example \cite[Eq.\ (3.7)]{Weniger/Steinborn/1983b}):
\begin{equation}
  \label{RBF_HalfInt}
  \hat{k}_{n+1/2} (z) \; = \; 2^n \, (1/2)_n \,
   \mathrm{e}^{-z} \, {}_1 F_1 (-n; -2n; 2z) \, .  
\end{equation}
A condensed review of the history of $B$ functions including numerous
references can be found in \cite{Weniger/2009a}. Reduced Bessel and $B$
functions had been the topic of my Diploma \citep{Weniger/1977} and my
PhD thesis \citep{Weniger/1982}.

\Crefrange{Def:B_Fun}{RBF_HalfInt} indicate that $B$ functions are fairly
complicated mathematical objects. Therefore, it is not at all obvious why
$B$ functions should offer any advantages. However, the Hankel-type
integral \citep[Eq.\ (2) on p.\ 410]{Watson/1922})
\begin{align}
 \label{B_Fun}
  & \int_{0}^{\infty} \, K_{\mu} (\alpha t) \, J_{\nu} (\beta t) \,
   t^{\mu+\nu+1} \, \mathrm{d} t
  \notag
  \\
  & \qquad \; = \; \Gamma (\mu+\nu+1) \,
    \frac{2^{\mu+\nu} \, \alpha^{\mu} \, \beta^{\nu}}
     {[\alpha^{2} + \beta^{2}]^{\mu+\nu+1}} \, , 
  \notag 
  \\
  & \qquad \qquad \Re (\mu+\nu) > \vert \Re (\mu) \vert \, , \quad
   \Re (\alpha) > \vert \Re (\beta) \vert \, ,
\end{align}
implies that a $B$ function possesses a Fourier transform of exceptional
simplicity:
\begin{align}
  \label{FT_B_Fun}
  & \overline{B_{n,\ell}^{m}} (\beta, \bm{p}) \; = \; (2\pi)^{-3/2} \, 
   \int \, \mathrm{e}^{- \mathrm{i} \bm{p} \cdot \bm{r}} \, 
    B_{n,\ell}^{m} (\beta, \bm{r}) \, \mathrm{d}^3 \bm{r}
  \notag 
  \\
  & \qquad \; = \; 
    (2/\pi)^{1/2} \, \frac{\beta^{2n+\ell-1}}{[\beta^{2} +
    p^{2}]^{n+\ell+1}} \, \mathcal{Y}_{\ell}^{m} (- \mathrm{i} \bm{p}) \, .
\end{align}
This is the most consequential and also the most often cited result of my
PhD thesis \cite[Eq.\ (7.1-6) on p.\ 160]{Weniger/1982}. Later, the
Fourier transform \eqref{FT_B_Fun} was published in \citep*[Eq.\
(3.7)]{Weniger/Steinborn/1983a}. Independently and almost simultaneously,
\cref{FT_B_Fun} was also derived by \citet[Eqs.\ (57) -
(58)]{Niukkanen/1984c}.

It follows from \cref{RBF_HalfInt} that a $B$ function can be expressed
as a finite sum of Slater-type functions, or equivalently, that the
Fourier transform \eqref{FT_B_Fun} of a $B$ function can be expressed as
a linear combination of the Fourier transforms of Slater-type functions,
just as \citet*{Yuekcue/Yuekcue/2018} had done it in the case of
bound-state hydrogen eigenfunctions (compare
\cref{Sec:WorkYuekcueYuekcue}).

\citet{Yuekcue/2015} used this seemingly simple approach of expressing a
$B$ function as a linear combination of Slater-type functions \citep[Eq.\
(21)]{Yuekcue/2015}. For the Fourier transform of a Slater-type function
-- his Eqs.\ (32), (39), and (40) -- he used the same expressions as the
ones used by \citet*[Eqs. (13) - (16)]{Yuekcue/Yuekcue/2018}. This leads
to explicit expressions \citep[Eqs.\ (41) and (42)]{Yuekcue/2015} that
are, however, much more complicated and therefore much less useful than
the remarkably compact Fourier transform \eqref{FT_B_Fun}.

We do not know for sure whether \citet*{Yuekcue/Yuekcue/2018} were aware
of the Fourier transform \eqref{FT_HydEigFun} derived by \citet*[Eq.\
(28)]{Podolsky/Pauling/1929} or of the other earlier references mentioned
in \cref{Sec:Intro}. Maybe, \citeauthor{Yuekcue/Yuekcue/2018} genuinely
believed that their results for the Fourier transform of a bound-state
hydrogen eigenfunctions are actually the best possible. However,
\citet{Yuekcue/2015} did not only present his fairly complicated Eqs.\
(41) and (42) for the Fourier transform of a $B$ function, but as his
Eq.\ (28) also the very compact expression \eqref{FT_B_Fun}. It is hard
to imagine that anyone would want to use \citeauthor{Yuekcue/2015}'s
complicated Eqs.\ (41) and (42) instead of the much simpler
\cref{FT_B_Fun}. Not all expressions, which are mathematically correct,
are useful and deserve to be published.

The exceptionally simple Fourier transform \eqref{FT_B_Fun} gives $B$
functions a special position among exponentially decaying functions. It
explains why other exponentially decaying functions as for example
Slater-type functions with integral principal quantum numbers, bound
state hydrogen eigenfunctions, and other functions based on generalized
Laguerre polynomials can be expressed in terms of finite linear
combinations of $B$ functions (for details, see \cite[Section
IV]{Weniger/1985} or \cite[Section 4]{Weniger/2002}).

The Fourier transform \eqref{FT_B_Fun} was extensively used by Safouhi
and co-workers for the evaluation of molecular multicenter integrals with
the help of numerical quadrature combined with extrapolation
techniques. Many references of the Safouhi group can be found in the PhD
thesis of \citet{Slevinsky/2014}.

Apart from the Fourier transform \eqref{FT_B_Fun}, the most important
expression of this Section is the expansion of a generalized Laguerre
polynomial in terms of reduced Bessel functions with half-integral
indices \citep[Eq.\ (3.3-35) on p.\ 45]{Weniger/1982}:
\begin{align}
  \label{GLag_FinSum_RBF}
  & \mathrm{e}^{-z} \, L_{n}^{(\alpha)} (2z) \; = \; (2n+\alpha+1) 
  \notag
  \\
  & \quad \times \,
   \sum_{\nu=0}^{n} \, \frac{(-2)^{\nu} \Gamma (n+\alpha+\nu+1)}
    {\nu! (n-\nu)! \Gamma (\alpha+2\nu+2)} \, \hat{k}_{\nu+1/2} (z) \, .
\end{align}
This relationship was used by \citet[Eq.\ (3.17)]{Filter/Steinborn/1980}
for the construction of addition theorems and other expansions in terms
of Lambda functions.

With the help of \cref{GLag_FinSum_RBF}, it is trivially simple to
express Sturmians and Lambda functions as finite linear combinations of
$B$ functions \cite[Eqs.\ (4.19) and (4.20)]{Weniger/1985}:
\begin{widetext}
  \begin{align}
    \label{Sturm_Bfun}
    \Psi_{n, \ell}^{m} (\beta, \bm{r}) & \; = \;
     \frac{(2 \beta)^{3/2} \, 2^{\ell}}{(2\ell+1)!!} \,
      \left[ \frac{2n(n+\ell)!}{(n-\ell-1)!} \right]^{1/2} \,
       \sum_{\nu=0}^{n-\ell-1} \, \frac{(-n+\ell+1)_{\nu} \, 
        (n+\ell+1)_{\nu}}{\nu! \, (\ell+3/2)_{\nu}} \,
         B_{\nu+1,\ell}^{m} (\beta, \hm{r}) \, ,
    \\
    \label{Lambda_Bfun}
    \Lambda_{n, \ell}^{m} (\beta, \bm{r}) & \; = \;
     (2 \beta)^{3/2} \, 2^{\ell} \, \frac{(2n+1)}{(2\ell+3)!!} \,
      \left[ \frac{(n+\ell+1)!}{(n-\ell-1)!} \right]^{1/2} \,
       \sum_{\nu=0}^{n-\ell-1} \, \frac{(-n+\ell+1)_{\nu} \,
        (n+\ell+2)_{\nu}}{\nu! \, (\ell+5/2)_{\nu}} \,
         B_{\nu+1,\ell}^{m} (\beta, \hm{r}) \, .
  \end{align} 
Now, we only need the Fourier transform (\ref{FT_B_Fun}) of a $B$
function to obtain explicit expressions for the Fourier transforms of a
Sturmian or of a Lambda function. By combining
\cref{FT_B_Fun,Sturm_Bfun,Lambda_Bfun}, we obtain the following
hypergeometric representations:
\begin{align}
  \label{FT_Sturm_2F1}
  & \overline{\Psi_{n, \ell}^{m}} (\beta, \bm{p}) \; = \;
   (2\pi)^{-3/2} \, \int \, \mathrm{e}^{-\mathrm{i} \bm{p} \cdot \bm{r}}
    \, \Psi_{n, \ell}^{m} (\beta, \bm{r}) \, \mathrm{d}^{3} \bm{r}
  \notag
  \\
  & \quad \; = \; \frac{1}{(2\ell+1)!!} \, \left[ \frac{\beta}{\pi} \, 
   \frac {2n \, (n+\ell)!} {(n-\ell-1)!} \right]^{1/2} \, \left[
    \frac{2\beta}{\beta^2+p^2} \right]^{\ell+2} \, 
     \mathcal{Y}_{\ell}^{m} (- \mathrm{i} \bm{p}) \,
      {}_2 F_1 \left( -n+\ell+1, n+\ell+1;
    \ell+\frac{3}{2}; \frac{\beta^2}{\beta^2+p^2} \right) \, ,
  \\
  \label{FT_Lambda_2F1}
  & \overline{\Lambda_{n, \ell}^{m}} (\beta, \bm{p}) \; = \; (2\pi)^{-3/2} \,
  \int \, \mathrm{e}^{-\mathrm{i} \bm{p} \cdot \bm{r}} \, \Lambda_{n,
    \ell}^{m} (\beta, \bm{r}) \, \mathrm{d}^3 \bm{r}
  \notag 
  \\
  & \quad \; = \; \frac{(2n+1)}{(2\ell+3)!!} \, \left[ \frac{\beta \,
   (n+\ell+1)!}{\pi \, (n-\ell-1)!} \right]^{1/2} \, \left[
    \frac{2\beta}{\beta^2+p^2} \right]^{\ell+2} \, 
    \mathcal{Y}_{\ell}^{m} (- \mathrm{i} \bm{p}) \,
     {}_2 F_1 \left( -n+\ell+1, n+\ell+2;
      \ell+\frac{5}{2}; \frac{\beta^2}{\beta^2+p^2} \right) \, .  
\end{align}
The terminating ${}_{2} F_{1}$ in \cref{FT_Sturm_2F1} can according to
\cref{GegPol_2F1_a} be replaced as a Gegenbauer polynomial, yielding
\cref{FT_SturmFun} \cite[Eq.\ (4.24)]{Weniger/1985}, and the terminating
${}_{2} F_{1}$ in \cref{FT_Lambda_2F1} can be expressed as a Jacobi
polynomial \citep*[p.\ 212]{Magnus/Oberhettinger/Soni/1966} via
\begin{equation}
  \label{JacPol_2F1_a}
  P_{n}^{(\alpha, \beta)} (x) \; = \; \binom{n+\alpha}{n} \,
  {}_2 F_1 \left( -n, \alpha+\beta+n+1; \alpha+1;
   \frac{1-x}{2} \right) \, ,  
\end{equation}
yielding the following explicit expressions for the Fourier transforms of
a Lambda function \citep[Eq.\ (4.25)]{Weniger/1985}:
\begin{align}
  \label{FT_Lambda}
  \overline{\Lambda_{n, \ell}^{m}} (\beta, \bm{p}) & \; = \;
   \frac{2}{(1/2)_n} \, \left\{ \frac{\beta \, 
    (n+\ell+1)! \, (n-\ell-1)!}{\pi} \right\}^{1/2} \, \left[
    \frac{\beta}{\beta^2+p^2} \right]^{\ell+2} \,
     \mathcal{Y}_{\ell}^{m} (-\mathrm{i} \bm{p}) \,
      P_{n-\ell-1}^{(\ell+3/2, \ell+1/2)} 
       \left( \frac{p^2-\beta^2}{p^2+\beta^2} \right) \, .
\end{align}
\end{widetext}

The orthogonality relationships satisfied by the Fourier transforms of
Sturmians and Lambda functions with respect to an integration over the
whole three-dimensional momentum space can be deduced directly from the
known orthogonality properties of the Gegenbauer and Jacobi polynomials
\citep[Eqs.\ (4.31) - (4.37)]{Weniger/1985}.

My approach, which is based on the \cref{GLag_FinSum_RBF,FT_B_Fun}, can
also be employed in the case of other, more complicated exponentially
decaying functions. In \citep[Abstract or Eqs.\ (1) and
(2)]{Guseinov/2002c}, \citeauthor{Guseinov/2002c} introduced a large
class of complete and orthonormal functions. In terms of the polynomials
$\bigl[ L_{q}^{{p}} (x) \bigr]_{\text{BS}}$ defined in
\cref{AssLagFun_BS}, \citeauthor{Guseinov/2002c}'s functions can be
expressed as follows:
\begin{align}
  \label{OrigDef_Psi_Guseinov}
  & \Psi_{n \ell m}^{\alpha} (\zeta, \bm{r}) \; = \;
   (-1)^{\alpha} \, \left[ \frac{(2 \zeta)^{3} (n-\ell-1)!}
    {(2n)^{\alpha} (n+\ell+1-\alpha)!} \right]^{1/2} \, 
  \notag
  \\
  & \quad \times \, (2 \zeta r)^{\ell} \, \mathrm{e}^{-\zeta r} \, 
  \bigr[ L_{n+\ell+1-\alpha}^{2\ell+2-\alpha} \bigl]_{\text{BS}}
  (2\zeta r) \, S_{\ell m} (\theta, \varphi) \, .
\end{align}
Here, $\zeta > 0$ is a scaling parameter, and
$S_{\ell m} (\theta, \varphi)$ is either a real or a complex spherical
harmonic (Guseinov did not provide an exact definition of
$S_{\ell m} (\theta, \varphi)$). 

The additional parameter $\alpha$, which Guseinov calls \emph{frictional}
or \emph{self-frictional quantum number}, was originally chosen to be an
integer satisfying $\alpha = 1, 0, -1, -2, \cdots$
\citep[Abstract]{Guseinov/2002c}. In the text following \citep[Eq.\
(3)]{Guseinov/2002c}, \citeauthor{Guseinov/2002c} remarked that for
\emph{fixed} $\alpha = 1, 0, -1, -2, \cdots$ the functions
\eqref{OrigDef_Psi_Guseinov} form a \emph{complete orthonormal set}.

This statement is meaningless. Completeness is not a generally valid
property of a given function set. It only guarantees that functions
belonging to a suitable Hilbert space, which has to be specified, can be
expanded by this function set, and that the resulting expansions converge
with respect to the norm of this Hilbert space (for further details, I
recommend a book by \citet{Higgins/1977} or a review by
\citet{Klahn/1981}).

\citeauthor{Guseinov/2002c}'s original definition
\eqref{OrigDef_Psi_Guseinov} implies that his functions are according to
\citep[Eq.\ (4)]{Guseinov/2002c} orthogonal with respect to the weight
function $w (r) = [n'/(\zeta r)]^{\alpha}$ \citep[Eq.\
(4)]{Guseinov/2002c}:
\begin{align}
  \label{Orig_Psi_Guseinov_Orthogon}
  & \int \, \left[ \Psi_{n \ell m}^{\alpha} (\zeta, \bm{r}) \right]^{*}
   \, \left( \frac{n'}{\zeta r} \right)^{\alpha} \,  
    \Psi_{n' \ell' m'}^{\alpha} (\zeta, \bm{r}) \, \mathrm{d}^{3} \bm{r}
  \notag
  \\
  & \qquad \; = \; 
  \delta_{n n'} \, \delta_{\ell \ell'} \, \delta_{m m'} \, .  
\end{align}
In the theory of classical orthogonal polynomials, which is intimately
linked to Hilbert space theory, it is common practice to introduce on the
basis of their orthogonality relationships suitable inner products
$( f \vert g )_{w} = \int_{a}^{b} w (x) [f (x)]^{*} g (x) \mathrm{d} x$
with a positive weight function $w \colon [a, b] \to
\mathbb{R}_{+}$. These weighted inner products then lead to the
corresponding weighted Hilbert spaces $\mathcal{H}_{w}$ in which the
orthogonal polynomials under consideration are complete and orthogonal.

In the case of Guseinov's orthogonality relationship
\eqref{Orig_Psi_Guseinov_Orthogon}, this approach does not work. The
weight function $w (r) = [n'/(\zeta r)]^{\alpha}$ cannot be used to
define a Hilbert space because both $\zeta$ and $n'$ are in general
undefined. Thus, instead of incorporating $\zeta$ and $n'$ into the
weight function, they should be incorporated in the normalization factor.

A further disadvantage of Guseinov's original definition
\eqref{OrigDef_Psi_Guseinov} is its use of the polynomials
$\bigl[ L_{q}^{{p}} (x) \bigr]_{\text{BS}}$ defined by
\cref{AssLagFun_BS}, which can only have integral superscripts. As an
alternative, I suggested the following definition, which uses the modern
mathematical notation for the generalized Laguerre polynomials (see for
example \citep[Eq.\ (4.16)]{Weniger/2007b} or \citep[Eq.\
(2.13)]{Weniger/2012}):
\begin{align}
  \label{Def_Psi_Guseinov}
  & \prescript{}{k}{\Psi}_{n, \ell}^{m} (\beta, \bm{r}) \; = \; \left[
   \frac{(2\beta)^{k+3} (n-\ell-1)!}{\Gamma (n+\ell+k+2)} \right]^{1/2}
  \notag
  \\
  & \qquad \times \, \mathrm{e}^{-\beta r} \,
   L_{n-\ell-1}^{(2\ell+k+2)} (2 \beta r) \,
    \mathcal{Y}_{\ell}^{m} (2 \beta \bm{r}) \, . 
\end{align}
The indices satisfy $n \in \mathbb{N}$,
$\ell \in \mathbb{N}_0 \le n - 1$, $-\ell \le m \le \ell$, and the
scaling parameter satisfies $\beta > 0$.

In my original definition in \citep[Eq.\ (4.16)]{Weniger/2007b} or
\citep[Eq.\ (2.13)]{Weniger/2012}, I had assumed that $k$ is a positive
or negative integer satisfying $k = -1, 0, 1, 2, \dots$, which
corresponds to the straightforward translation $-\alpha \mapsto k$ of
Guseinov's original condition $\alpha = 1, 0, -1, -2, \dots$ \citep[Eq.\
(4)]{Guseinov/2002c}. Therefore, my original definition in \citep[Eq.\
(4.16)]{Weniger/2007b} or \citep[Eq.\ (2.13)]{Weniger/2012} assumed $k$
being integral and contained $(n+\ell+k+1)!$ instead of
$\Gamma (n+\ell+k+2)$.

However, in the text following \citep[Eq.\ (4.16) on p.\
11]{Weniger/2007b} or in the text following \citep[Eq.\ (2.13) on p.\
27]{Weniger/2012}, I had emphasized that the condition
$k = -1, 0, 1, 2, \dots$ is unnecessarily restrictive and that it can be
generalized to $k \in [-1, \infty)$. My criticism of Guseinov's original
definition \eqref{OrigDef_Psi_Guseinov} was implicitly confirmed by
Guseinov himself. In his later articles
\citep{Guseinov/2012a,Guseinov/2013a}, Guseinov generalized his so-called
frictional quantum number from originally $\alpha = 1, 0, -1, -2, \cdots$
to $\alpha \in (-\infty, 3)$, which corresponds to $k \in (-3, \infty)$
in my notation. This change could not be done with Guseinov's original
definition \eqref{OrigDef_Psi_Guseinov}.

Therefore, Guseinov finally had to use the modern mathematical notation
for his functions (compare \citep[Eqs.\ (1) - (5)]{Guseinov/2012a} or
\citep[Abstract]{Guseinov/2013a}). To disguise the obvious, Guseinov used
in these formulas instead of a generalized Laguerre polynomial a
terminating confluent hypergeometric series ${}_{1} F_{1}$. Because of
\cref{GLag_1F1}, Guseinov's formulas are equivalent to my definition
\eqref{Def_Psi_Guseinov}. Characteristically, Guseinov did not
acknowledge my contributions \citep[Eq.\ (4.16)]{Weniger/2007b} or
\citep[Eq.\ (2.13)]{Weniger/2012} to his functions. Because of
\citep{Guseinov/2007a}, Guseinov cannot claim to be unaware of
\citep{Weniger/2007b}.

For fixed $k \in (-3, \infty)$, Guseinov's functions defined by
\cref{Def_Psi_Guseinov} satisfy the orthonormality relationship
\begin{align}
  \label{Psi_Guseinov_OrthoNor}
  & \int \, \bigl[ \prescript{}{k}{\Psi}_{n, \ell}^{m} 
   (\beta, \bm{r}) \bigr]^{*} \, r^{k} \, 
    \prescript{}{k}{\Psi}_{n', \ell'}^{m'} 
    (\beta, \bm{r}) \, \mathrm{d}^3 \bm{r}
  \notag
  \\
  & \qquad \; = \; 
  \delta_{n n'} \, \delta_{\ell \ell'} \, \delta_{m m'} \, ,
\end{align}
which implies that they are complete and orthonormal in the weighted
Hilbert space
\begin{align}
  \label{HilbertL_r^k^2}
  & L_{r^k}^{2} (\mathbb{R}^3)
  \notag
  \\
  & \qquad \; = \; \Bigl\{ f \colon \mathbb{R}^3 \to
   \mathbb{C} \Bigm\vert \, \int \, r^k \, \vert f (\bm{r}) \vert^2 \,
    \mathrm{d}^3 \bm{r} < \infty \Bigr\} \, .
\end{align}

For $k=0$, the functions
$\prescript{}{k}{\Psi}_{n, \ell}^{m} (\beta, \bm{r})$ are identical to
the Lambda functions defined by \cref{Def_LambdaFun}. Thus, they are
complete and orthonormal in the Hilbert space $L^{2} (\mathbb{R}^3)$ of
square integrable functions defined by \cref{HilbertL^2}.

For $k = - 1$, Guseinov's functions yield apart from a different
normalization the Sturmians \eqref{Def_SturmFun}, which are complete and
orthonormal in the Sobolev space $W_{2}^{(1)} (\mathbb{R}^3)$, or
complete and orthogonal in the weighted Hilbert space
$L_{1/r}^{2} (\mathbb{R}^3)$.

Personally, I prefer Sturmians satisfying \cref{Def_SturmFun}:
$W_{2}^{(1)} (\mathbb{R}^3)$ is a proper subspace of
$L^{2} (\mathbb{R}^3)$ with some additional advantageous features
\cite{Klahn/1981}, whereas $L_{1/r}^{2} (\mathbb{R}^3)$ is not.

For $k \ne 0$, the weighted Hilbert spaces $L_{r^k}^{2} (\mathbb{R}^3)$
are genuinely different from the Hilbert space $L^{2} (\mathbb{R}^3)$ of
square integrable functions. We neither have
$L^{2} (\mathbb{R}^3) \subset L_{r^k}^{2} (\mathbb{R}^3)$ nor
$L_{r^k}^{2} (\mathbb{R}^3) \subset L^{2} (\mathbb{R}^3)$. In quantum
physics, it is tacitly assumed that bound-state wave functions are square
integrable \citep{Born/1955}. However, approximation processes converging
with respect to the norm of the weighted Hilbert space
$L_{r^k}^{2} (\mathbb{R}^3)$ with $k \ne -1, 0$ could produce functions
that are not square integrable. Obviously, this would lead to some
embarrassing conceptual and technical problems.

\begin{widetext}
With the help of \cref{GLag_FinSum_RBF}, Guseinov's functions can be
expressed as a finite sum of $B$ functions \citep[Eq.\
(2.22)]{Weniger/2012}:
\begin{align}
  \label{GusFun_Bfun}
  & \prescript{}{k}{\Psi}_{n, \ell}^{m} (\beta, \bm{r}) \; = \; \left\{
    \frac{\beta^{k+3} \, (n+\ell+k+1)!}{2^{k+1} \, (n-\ell-1)!}
    \right\}^{1/2}
    \frac{(2n+k+1) \, \Gamma (1/2) \, (\ell+1)!}  {\Gamma
    \bigl(\ell+2+k/2\bigr) \, \Gamma \bigl(\ell+[k+5]/2\bigr)}
  \notag \\
  & \qquad \times \sum_{\nu=0}^{n-\ell-1} \, \frac{(-n+\ell+1)_{\nu} \,
    (n+\ell+k+2)_{\nu} \, (\ell+2)_{\nu}}{\nu!  \,
    \bigl(\ell+2+k/2\bigr)_{\nu} \, \bigl(\ell+[k+5]/2\bigr)_{\nu}} \,
  B_{\nu+1,\ell}^{m} (\beta, \hm{r}) \, .
\end{align}
Now, we only need the Fourier transform (\ref{FT_B_Fun}) to obtain an
explicit expression for the Fourier transform of Guseinov's function:
\begin{align}
  \label{FT_GusFun}
  & \overline{\prescript{}{k}{\Psi}_{n, \ell}^{m}} (\beta, \bm{p})
   \; = \; (2\pi)^{-3/2} \, \int \,
    \mathrm{e}^{-\mathrm{i} \bm{p} \cdot \bm{r}} \,
    \prescript{}{k}{\Psi}_{n, \ell}^{m} (\beta, \bm{r}) \,
     \mathrm{d}^{3} \bm{r}
  \notag
  \\
  & \qquad \; = \; \left\{ \frac{\beta^{k+1} \, (n+\ell+k+1)!}
   {\pi \, 2^k \, (n-\ell-1)!} \right\}^{1/2} \, \frac
    {(2n+k+1) \, \Gamma (1/2) \, (\ell+1)!}
     {2^{\ell+2} \, \Gamma \bigl(\ell+2+k/2\bigr) \,
    \Gamma \bigl(\ell+[k+5]/2\bigr)} \,
     \mathcal{Y}_{\ell}^{m} (- \mathrm{i} \bm{p})
  \notag
  \\
  & \qquad \qquad \times \, \left( \frac{2\beta}{\beta^2+p^2}
   \right)^{\ell+2} \, {}_3 F_2 \biggl( -n+\ell+1, n+\ell+2, \ell+2;
    \ell+2+\frac{k}{2}, \ell+\frac{k+5}{2};
     \frac{\beta^2}{\beta^2+p^2} \biggr) \, .
\end{align}   
In this Fourier transform, the radial part is essentially a terminating
generalized hypergeometric series ${}_3 F_2$, which simplifies for either
$k = -1$ or $k = 0$ to yield the terminating Gaussian hypergeometric
series ${}_{2} F_{1}$ in the hypergeometric representations
\eqref{FT_Sturm_2F1} or \eqref{FT_Lambda_2F1} for the Fourier transforms
of Sturmians and Lambda functions, respectively. Thus, the Fourier
transform of Guseinov's function
$\prescript{}{k}{\Psi}_{n, \ell}^{m} (\beta, \bm{r})$ with $k \ne -1, 0$
is more complicated than the Fourier transforms of either Sturmians or
Lambda functions.
\end{widetext}

%
\typeout{==> Section: Summary and Conclusions}
\section{Summary and Conclusions}
\label{Sec:SummaryAndConclusions}
%

The Fourier transform \eqref{FT_HydEigFun} for a bound-state hydrogen
eigenfunction \eqref{Def_HydEigFun} is a classic result of quantum
physics already derived in \citeyear{Podolsky/Pauling/1929} by
\citet*[Eq.\ (28)]{Podolsky/Pauling/1929} with the help of the generating
function \eqref{GLag_GenFun}. I am not aware of a more compact and more
useful expression for this Fourier transform.

\citet*{Yuekcue/Yuekcue/2018}, who were apparently unaware not only of
\citeauthor{Podolsky/Pauling/1929}, but also of the other references
listed in \cref{Sec:Intro}, proceeded differently. As discussed in
\cref{Sec:WorkYuekcueYuekcue}, \citet*{Yuekcue/Yuekcue/2018} expressed a
generalized Laguerre polynomial as a finite sum of powers according to
\cref{GLag->PowZ}, or equivalently, they expressed a bound-state hydrogen
eigenfunction as a finite sum of Slater-type functions. Since Fourier
transformation is a linear operation, this leads to an expression of the
Fourier transformation of a bound-state hydrogen eigenfunction as a
finite sum of Fourier transforms of Slater-type functions, for which many
explicit expressions are known in the literature (compare
\cref{Sec:WorkYuekcueYuekcue}).

At first sight, this approach, which requires no mathematical skills,
looks like a good idea. Unfortunately, the simplicity of Slater-type
functions in the coordinate representation is deceptive. As already
emphasized in \citep*[p.\ 29]{Weniger/2012}, the Fourier transforms of
bound-state hydrogen eigenfunctions and Slater-type function have the
same level of complexity. Consequently, it cannot be a good idea to
express the Fourier transform of a bound-state hydrogen eigenfunction as
a linear combination of Fourier transforms of Slater-type
functions. Moreover, in the case of large principal quantum numbers $n$,
these finite sums tend to become numerically unstable. This is a direct
consequence of the alternating signs in \cref{GLag->PowZ}.

In principle, it should be possible to derive the
\citeauthor{Podolsky/Pauling/1929} formula \eqref{FT_HydEigFun} from the
comparatively complicated linear combinations presented by
\citeauthor{Yuekcue/Yuekcue/2018}. However, the Fourier transforms of
Slater-type functions discussed in \cref{Sec:WorkYuekcueYuekcue} are all
fairly complicated objects. Therefore, it is very difficult or even
practically impossible to obtain the remarkably compact
\citeauthor{Podolsky/Pauling/1929} formula \eqref{FT_HydEigFun} in this
way. I am not aware of anybody who achieved this.

It is nevertheless possible to derive the
\citeauthor{Podolsky/Pauling/1929} formula \eqref{FT_HydEigFun} by
expanding generalized Laguerre polynomials, albeit in terms of some
other, less well known polynomials. This was shown in
\cref{Sec:ExpRBF}. The key relationships in \cref{Sec:ExpRBF} are the
exceptionally simple Fourier transform \eqref{FT_B_Fun} of a $B$ function
and the expansion \eqref{GLag_FinSum_RBF} of a generalized Laguerre
polynomial in terms of reduced Bessel functions \eqref{RBF_HalfInt} with
half-integral indices. With the help of \cref{FT_B_Fun,GLag_FinSum_RBF}
it is trivially simple to derive the \citeauthor{Podolsky/Pauling/1929}
formula \eqref{FT_HydEigFun}. This derivation is much simpler than the
original derivation by \citet*{Podolsky/Pauling/1929}, which is discussed
in \cref{Sec:PodolskyPauling} and which required the skillful use of the
generating function \eqref{GLag_GenFun} of the generalized Laguerre
polynomials.

\citet{Hylleraas/1928} observed already in \citeyear{Hylleraas/1928} that
bound-state hydrogen eigenfunctions \emph{without} the inclusion of the
mathematically very difficult continuum eigenfunctions are incomplete in
the Hilbert space of square integrable functions (compare
\cref{Sec:Incom_BoundStateHydrEigFun}). This is a highly consequential
fact, which \citeauthor{Yuekcue/Yuekcue/2018} were apparently not aware
of. In combination with the difficult nature of the continuum
eigenfunctions, this incompleteness greatly limits the practical
usefulness of bound-state eigenfunctions as mathematical tools. It is
certainly not a good idea to do expansions in terms of an incomplete
function set.

Therefore, attention has shifted away from hydrogen eigenfunctions to
other, related function sets also based on the generalized Laguerre
polynomials, which, however, have more convenient completeness
properties. The best known examples are the so-called Sturmians
\eqref{Def_SturmFun}, which had been introduced by \citet{Hylleraas/1928}
already in \citeyear{Hylleraas/1928} and which can be obtained from the
bound-state hydrogen eigenfunctions by the substitution
$Z/n \mapsto \beta$ according to \cref{SturmFun<->BSHEF}, and the
so-called Lambda functions \eqref{Def_LambdaFun}, which were also
introduced by \citet{Hylleraas/1929} in \citeyear{Hylleraas/1929}.

With the help of the expansion \eqref{GLag_FinSum_RBF}, it is a trivial
matter to express both Sturmians and Lambda functions as linear
combinations of $B$ functions, yielding
\cref{Sturm_Bfun,Lambda_Bfun}. Then, one only needs the Fourier transform
\eqref{FT_B_Fun} to convert these linear combinations to compact explicit
expressions for the Fourier transforms of Sturmians and Lambda functions,
respectively.

In \citeyear{Guseinov/2002c}, \citet{Guseinov/2002c} introduced a large
class of complete and orthonormal functions defined by
\cref{OrigDef_Psi_Guseinov}, which used an antiquated notation for the
Laguerre polynomials. Guseinov's functions contain an additional
parameter $\alpha$ called \emph{frictional} or \emph{self-frictional
  quantum number}, which was originally assumed to be integral. Depending
on this $\alpha$, Guseinov's functions contain Sturmians and Lambda
functions as special cases.

Guseinov's original notation \eqref{OrigDef_Psi_Guseinov} does not allow
non-integral values of $\alpha$.  In order to rectify this obvious
deficiency, I introduced in \citep{Weniger/2007b,Weniger/2012} the
alternative definition \eqref{Def_Psi_Guseinov} which uses the modern
mathematical notation for the generalized Laguerre polynomials. Later,
\citet{Guseinov/2012a,Guseinov/2013a} was forced to change to my notation
because he wanted to consider non-integral self-frictional quantum
numbers $\alpha$.

With the help of \cref{FT_B_Fun,GLag_FinSum_RBF}, it is again a trivial
matter to construct the Fourier transform \eqref{FT_GusFun} of a Guseinov
function in the notation of \cref{Def_Psi_Guseinov}. \Cref{FT_GusFun}
contains the Fourier transforms of Sturmians and Lambda functions as
special cases, but is more complicated.

%
%

%

\end{document}